\documentclass[12pt,a4paper,final]{iopart}
\usepackage{iopams,braket}
\usepackage{graphicx,overpic,floatrow}
\usepackage[caption=false]{subfig}
\usepackage{color, colortbl}
\definecolor{Gray}{gray}{0.9}
\usepackage{soul}

\usepackage[colorlinks=true,citecolor=blue,linkcolor=black]{hyperref}

\begin{document}

\title{Cavity magnomechanical storage and retrieval of quantum states}
 
\author{Bijita Sarma, Thomas Busch}
\address{Quantum Systems Unit, Okinawa Institute of Science and Technology Graduate University, Okinawa 904-0495, Japan}

\author{Jason Twamley}
\address{Quantum Machines Unit, Okinawa Institute of Science and Technology Graduate University, Okinawa 904-0495, Japan\\
	and\\
	Centre for Engineered Quantum Systems, Department of Physics and Astronomy,
Macquarie University, Sydney, New South Wales 2109, Australia}

\vspace{10pt}
\begin{indented}
\item[] \today
\end{indented}

\begin{abstract}
We show how a quantum state in a microwave cavity mode can be transferred to and stored in a phononic mode via an intermediate magnon mode in a magnomechanical system. For this we consider a ferrimagnetic yttrium iron garnet (YIG) sphere inserted in a microwave cavity, where the microwave and magnon modes are coupled via a magnetic-dipole interaction and the magnon and phonon modes in the YIG sphere are coupled via magnetostrictive forces. By modulating the cavity and magnon detunings and the driving of the magnon mode in time, a Stimulated Raman Adiabatic Passage (STIRAP)-like coherent transfer becomes possible between the cavity mode and the phonon mode. The phononic mode can be used to store the photonic quantum state for long periods as it possesses lower damping than the photonic and magnon modes. 
Thus our proposed scheme offers a possibility of using magnomechanical systems as quantum memory for photonic quantum information.  
\end{abstract}

\vspace{2pc}
\noindent{\it Keywords}: Cavity Magnomechanics, Quantum Information Storage, Stimulated Raman Adiabatic Passage

\section{Introduction}
The coupling between magnons in a ferrimagnetic material and phonons in a mechanical resonator has attracted wide attention in recent years due to its application in magnomechanical settings similar to cavity quantum electrodynamics and optomechanics. A mechanical resonator can be coupled to an optical cavity mode via radiation pressure interaction \cite{kippenberg2005analysis,arcizet2006radiation}, as well as to a microwave cavity mode via an electrostatic interaction \cite{teufel2011circuit,andrews2014bidirectional}. Thus such mechanical mirrors can be used as  transducers for the coherent transfer between optical and microwave fields \cite{wang2012using,tian2012adiabatic}, or as optical wavelength converters \cite{hill2012coherent}. One of the relatively new candidates in the picture is the magnon, a collective excitation of magnetization, which can be coupled to phonons via the magnetostrictive force \cite{Zhang2016a}. This type of coupling has better tunability as the magnon frequency can be controlled via an external magnetic field \cite{serga2010yig,lenk2011building,chumak2015magnon}. Magnons can in turn be strongly coupled to microwave cavity modes as well, particularly in the insulating magnetic material yttrium iron garnet (YIG). The Kittel mode \cite{Kittel1948b}, found in YIG spheres, can couple very strongly to cavity photons or to superconducting quantum circuits \cite{Tabuchi2015}, producing a hybrid system that can be efficiently used in quantum information processing \cite{Huebl2013,Tabuchi2014,Zhang2014,Goryachev2014,Bai2015,Zhang2015a,li2018magnon,li2019squeezed,li2021entangling,qi2020magnon}. This kind of magnon-cavity coupling can also give rise to bistable behavior as studied in~\cite{Wang2018}.

Combining the magnetostrictive coupling between magnons and phonons and the magnon-cavity coupling, a new kind of photon-magnon-phonon interaction can be realized with YIG spheres interacting with microwave cavities. This opens up new possibilities in quantum state engineering and control. 
Here we consider a photon-magnon-phonon coupled system which is capable of transferring cavity photons to mechanical motional phonons, enabling the storage of photonic quantum states for long durations, thanks to the lower damping rate of the phononic mode. Storage of quantum states and information in long-lasting modes is very important for quantum communication networks. Photons are the usual candidate for the flying qubits, whereas some other quantum system with a longer coherence time is necessary for the storage of quantum information in quantum repeaters and networks \cite{briegel1998quantum}. The storage of quantum states has, for example, been previously studied in atomic media, where the information is stored as spin excitations \cite{liu2001observation,phillips2001storage,hammerer2010quantum},  in optomechanical systems \cite{fiore2011storing,fiore2013optomechanical,kumar2019single}, in coupled optical waveguides, or in acoustic excitations \cite{yanik2004stopping,xu2007breaking,baba2008slow,zhu2007stored}. 

The system we consider consists of a YIG-sphere placed inside a microwave cavity where the magnon modes of the sphere couple with the deformation phonon modes via a magnetostrictive force, and also to the electromagnetic cavity modes via a magnetic dipole interaction. We drive the magnon mode with a microwave field such that the time-modulation of the amplitude of the drive on the magnon mode yields a {\em time-dependent} coupling strength between the magnon and the mechanics. Using additionally time-modulated resonance frequencies of the photonic and magnon modes, a STIRAP-like one-way transfer becomes possible \cite{bergmann2019roadmap,bergmann2015perspective}, which efficiently transfers quantum states from the cavity mode to the mechanical mode. The quanta can be stored in the mechanical resonator for some time and can then be extracted using a retrieval pulse. In this way quantum states can be stored for times longer than the cavity lifetime, due to the lower damping rate of the phononic mode. We show that our storage protocol can be applied to various kinds of quantum states. 

\section{Model}
\begin{figure}
\centering
\includegraphics[width=0.8\columnwidth]{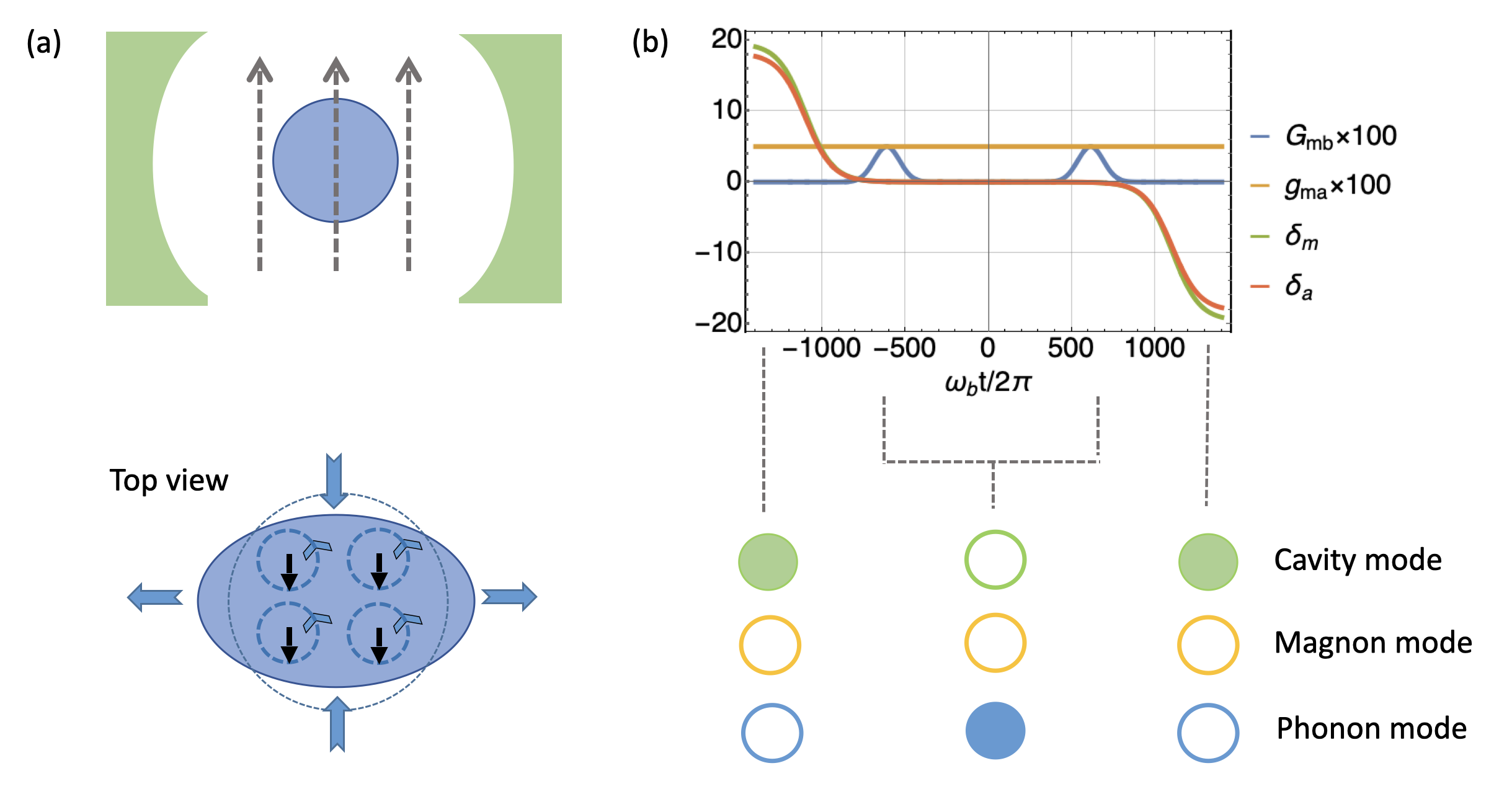}
\caption{(a) Schematic of the cavity magnomechanical system where a YIG-sphere is placed inside a microwave cavity so that the magnon modes of the sphere interact with the cavity mode via a collective-spin-photon coupling, and also with the deformation phonon modes via a magnetostrictive force. The vertical dashed arrows indicate an applied magnetic field and the dynamical magnetization causes a deformation of the YIG sphere (see top view in lower panel), 
which leads to magnetostrictive coupling between magnons and phonons \cite{Zhang2016a}.
(b) The coupling pulses and the detunings in units of $\omega_b/2\pi$, and the variation in the cavity, magnon and phonon states for different durations of the pulse sequence. This choice of pulses leads to a STIRAP-like one-way transfer for quantum states from the cavity mode $a$ to the mechanical mode $b$ and after some storage delay, a transfer back to the cavity mode. The state to be transferred is initially in the cavity mode and the filled circles indicate the mode that the quantum state is encoded in at particular times during the protocol. The parameters considered here are: $2\pi \Omega_0/\omega_b=0.1$, $\omega_b T = 108.7$, $\omega_b t_{c_1} = -612.2$, $\omega_b t_{c_2} = 612.2$, $\omega_b \tau_{ch} = 164.9$, $\omega_b \tau = 1101.6$, $\kappa_{\delta} = 14.05$, and $h_{\delta} = 13.94$. Here $\kappa_\delta$ and $h_\delta$ are dimensionless parameters. 
}
 \label{fig:schematic_model}
\end{figure}
We consider a system schematically shown in Fig.~\ref{fig:schematic_model}(a), where a YIG sphere is inserted into a microwave cavity. The Hamiltonian of the system is given by (note that we have set $\hbar =1$)
\begin{eqnarray}
H_0 &=& \omega _{a}a^{\dagger }a + \omega_m m^{\dagger}m + \omega_{b}b^{\dagger }b+ g_{ma}(a^{\dagger }m+m^{\dagger}a) + g_{mb}m^{\dagger}m(b+b^{\dagger })
\label{eq01} \nonumber\\
&& + i(\varepsilon_{p}m^{\dagger} e^{-i \omega_p t} - \varepsilon_{p}^{\ast}m e^{i \omega_p t}),  
\end{eqnarray}%
where $a(a^{\dagger})$, $m(m^{\dagger})$ and $b(b^{\dagger})$ are the annihilation (creation) operators of the microwave cavity mode, the magnon mode and the mechanical mode with resonance frequencies $\omega_a$, $\omega_m$ and $\omega_b$ respectively. The magnon frequency can be tuned by the
external bias magnetic field, $\mathcal H$, as $\omega_m = \gamma \mathcal H$, where $\gamma/2\pi=28\ \rm GHz/T$ is the gyromagnetic ratio. In addition, $g_{ma}$ and $g_{mb}$ are the single excitation coupling rates of the cavity-magnon interaction and magnon-phonon magnetostrictive interaction. 
The last term in (\ref{eq01}) describes the external driving of the magnon mode, where $\omega_p$ is the frequency of the drive magnetic field and $\varepsilon_{p}$ is the Rabi frequency between the drive magnetic field and the magnon mode, given by $\epsilon_p=\frac{\sqrt{5}}{4} \gamma \sqrt{N} B_0$~\cite{li2018magnon}. Here $B_0$ is the amplitude of the drive magnetic field, 
$N = \rho V$ is the total number of spins in a sphere of volume $V$ 
and $\rho = 4.22 \times 10^{27}\, {\rm m^{-3}}$ is the spin density of YIG. 

Moving over to the frame rotating with the drive frequency $\omega_p$, given by the transformation, $R =\exp   \big [i \omega_p\, (a^\dagger a  + m^\dagger m)\, t \big ]$, with $H = RH_0 R^\dagger + i \frac{\partial R}{\partial t}R^\dagger$, the Hamiltonian of the system can be written as 
\begin{eqnarray}
\nonumber
H &=&\Delta _{a}a^{\dagger }a + \Delta _{m}m^{\dagger}m + \omega
_{b}b^{\dagger }b+ g_{ma}(a^{\dagger }m+m^{\dagger}a)+g_{mb}m^{\dagger}m(b+b^{\dagger })
\label{eq02} \\
&&+i(\varepsilon _{p}m^{\dagger}-\varepsilon_{p}^{\ast }m),  
\end{eqnarray}%
where $\Delta _{a}=\omega _{a} - \omega _{p}$ and $\Delta _{m}=\omega _{m} - \omega _{p}$ are detunings. 

The dynamical evolution of the system operators can then be described by the Langevin equations 
\begin{eqnarray}
\dot{a} &=&(-i\Delta _{a}-\kappa _{a})a - ig_{ma}m + \sqrt{2\kappa _{a}}a_{\rm in},
\nonumber \\
\dot{b} &=&(-i\omega _{b}-\kappa _{b})b - ig_{mb}m^{\dagger}m+\sqrt{2\kappa
	_{b}}b_{\rm in},  \\
\dot{m} &=&(-i\Delta _{m}-\kappa _{m})m - ig_{ma}a-ig_{mb}m(b+b^{\dagger })
\label{eq2} +\varepsilon _{p}+\sqrt{2\kappa _{m}}m_{\rm in}.  \nonumber
\end{eqnarray}
Here $\kappa _{a},\kappa _{m}$ and $\kappa_{b}$ are the losses of
the cavity mode, the magnon mode and the mechanical mode respectively and $a_{\rm in}, m_{\rm in}$ and $b_{\rm in}$ are the noise operators
with zero mean values, and correlation functions given by $\left\langle x_{\rm in}(t)x_{\rm in}^{\dagger }(t^{\prime })\right\rangle =(\bar{n}_{x}+1)\delta (t-t^{\prime })$, and $\left\langle x_{\rm in}^{\dagger }(t)x_{\rm in}(t^{\prime })\right\rangle =\bar{n}_{x}\delta (t-t^{\prime })$, with $x=\{a, b, m\}$.
The $\bar{n}_{a}$, $\bar{n}_{m}$ and $\bar{n}_{b}$ are the mean thermal occupations of the cavity mode, magnon mode and phonon mode, given by $\bar{n}_{x}=(e^{\hbar \omega _{x}/k_{B}T}-1)^{-1}$, where $T$ is the
bath temperature and $k_{B}$ is the Boltzmann constant. For strong driving, each Heisenberg operator can be expressed as a sum of its steady-state mean value and the quantum fluctuation, i.e., $a=\alpha +a_1, b=\beta +b_1$ and $m=\eta +m_1$, where $\alpha$, $\beta$, $\eta$ are the classical mean field values of the modes and $a_1$, $b_1$, $m_1$ are the corresponding quantum fluctuation operators.

Following the standard linearization approach \cite{wang2012using,tian2012adiabatic}, the dynamics of the quantum fluctuations is given by the linearized Hamiltonian of the form
\begin{eqnarray}
H_{\rm lin} &=&\Delta _{a}a^{\dagger }_1 a_1+\tilde{\Delta}_{m}m^{\dagger}_1m_1+\omega
_{b}b^{\dagger }_1b_1 +G_{mb}(m_1+m^{\dagger}_1)(b_1+b^{\dagger }_1)  \nonumber \\
&&+g_{ma}(m^{\dagger}_1a_1+m_1a^{\dagger }_1),  
\end{eqnarray}%
where $G_{mb}=\eta g_{mb}$ is
the coherent-driving-enhanced linearized magnomechanical coupling strength, with $\eta = \varepsilon_p (i \Delta_a + \kappa_a)/(g_{ma}^2 + i (\tilde{\Delta}_m + \kappa_m)(i \Delta_a + \kappa_a))$. 
Here $\tilde{\Delta}_{m}=\Delta_m + g_{mb}(\beta + \beta^{*})$ is the magnomechanical interaction-induced effective magnon-drive
detuning. Transforming this Hamiltonian into an interaction picture with $U= \exp\left[-i\omega_b (a^\dagger_1 a_1 + m^\dagger_1 m_1 + b^\dagger_1 b_1)t\right]$ yields  
\begin{eqnarray}
H &=&\delta _{a} a^{\dagger }_1a_1+\delta_{m} m^{\dagger}_1m_1 +G_{mb}\bigg (m^{\dagger}_1b_1+m_1 b^{\dagger }_1 + e^{-2i\omega_b t} m_1 b_1+  \nonumber \\
&& e^{2i\omega_b t} m^\dagger_1 b^\dagger_1 \bigg )  
+g_{ma}(m^{\dagger}_1a_1+m_1a^{\dagger }_1).   \label{RWA1}
\end{eqnarray}%
Here $\delta_a = \Delta_a - \omega_b$ and $\delta_m = \tilde{\Delta}_m - \omega_b$ are the effective detunings. 
The magnomechanical coupling, $G_{mb}$, can be varied by tuning the magnon drive Rabi frequency, and 
the magnon frequency can be altered by adjusting the strength of the external magnetic bias field \cite{li2018magnon}, which can therefore be used to tune the magnon detuning, $\delta_m$. The photon frequency can be modulated by using a tunable 3D microwave cavity as realized in the Refs.~\cite{carvalho2014piezoelectric,c2016piezoelectric,clark2018cryogenic,ramp2020wavelength}, which will modulate the cavity detuning, $\delta_a$. 
In the following section we will show that, using these time-dependent modulations, a STIRAP-like protocol can be designed to effectively transfer the microwave cavity state to the mechanical mode and then retrieve it back to the microwave cavity mode.

\section{State transfer and retrieval}
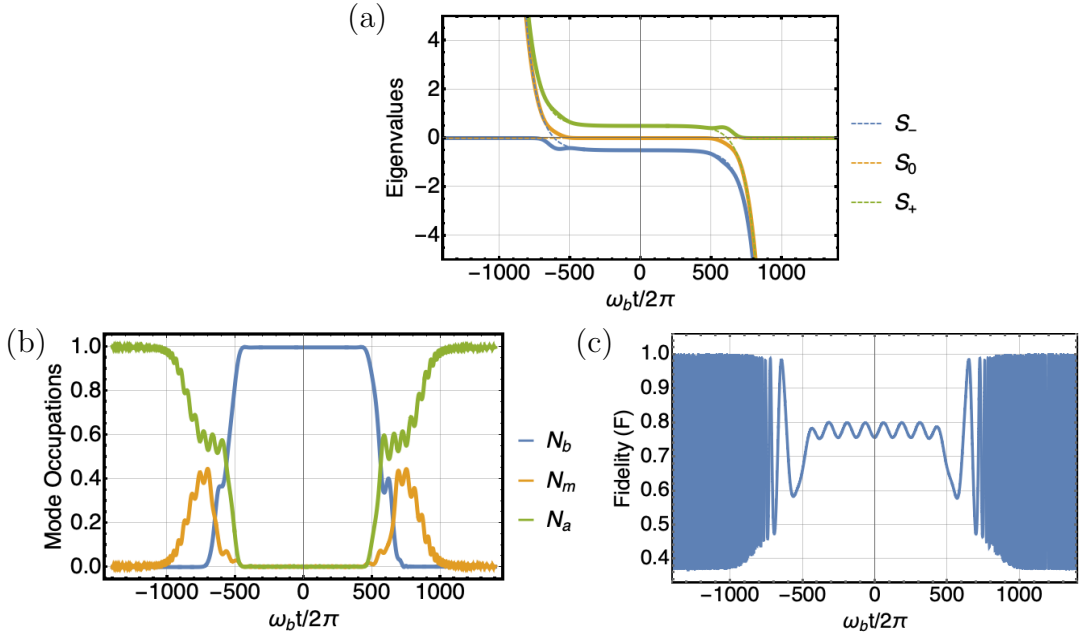
\begin{figure}[t] 
\begin{center}
\setlength{\unitlength}{1cm}
\begin{picture}(8.5,9)
\put(2,4){\includegraphics[width=.45\columnwidth]{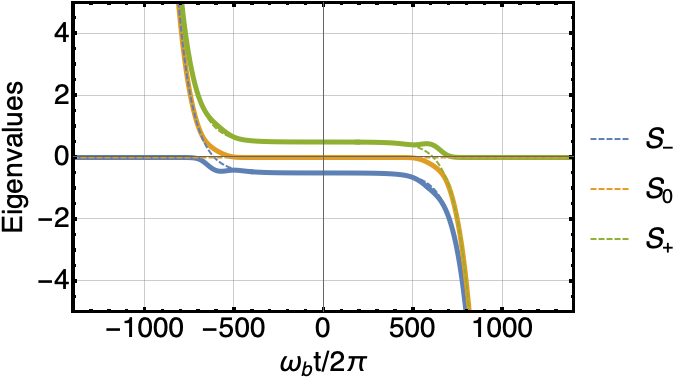}}
\put(1.5,7.8){(a)}
\put(-2.5,-.25){\includegraphics[width=.45\columnwidth]{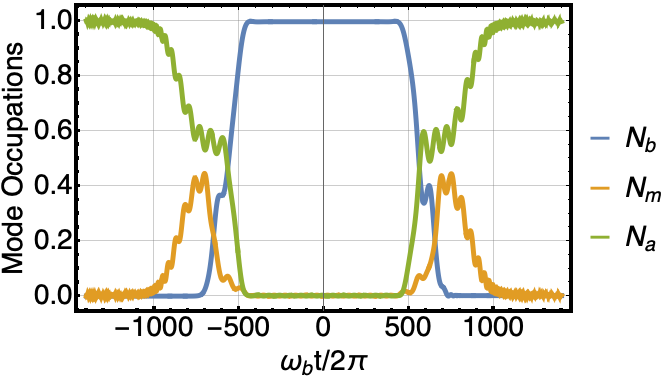}}
\put(-3,3.5){(b)}
\put(5,-.25){\includegraphics[width=.39\columnwidth]{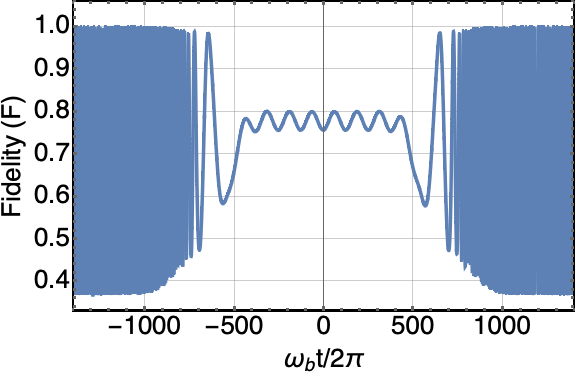}}
\put(4.5,3.5){(c)}
\end{picture}
\end{center}
\caption{(a) Evolution of the `Stokes' eigenvalues, $(S_0, S_+, S_-)$ (dashed lines), and the instantaneous eigenvalues of the system under the STIRAP-like pulses explained in the text (solid lines), (b) Unitary population transfer dynamics (without any damping) of the Fock state $|1,0,0\rangle$ showing the transfer from the microwave cavity mode to the mechanical resonator mode and then back to the cavity mode, (c) Unitary transfer dynamics of a coherent state with $\alpha=0.5$ in terms of the fidelity, $F$, from the microwave cavity mode to the mechanical resonator mode and then back to the cavity mode. 
The pulse parameters considered are same as in Fig.~\ref{fig:schematic_model}.
}
\label{fig:lossless_transfer}
\end{figure}

Stimulated Raman Adiabatic Passage - or STIRAP for short - can be efficiently used to transfer population in conventional three-level atomic systems \cite{bergmann2015perspective,vitanov2017stimulated}. 
Conventional STIRAP relies on the fact that 
at two-photon resonance an instantaneous eigenvector 
with zero eigenvalue exists,  
which is called a \textit{dark-state}, and which is a superposition of the initial and target states.
In the adiabatic limit the STIRAP dynamics allows to trap the population within the `dark state manifold' at all times, and any population transfer to the intermediate state, 
which has often a high decay rate, is avoided.  
The standard STIRAP protocol then modulates the coupling strengths in time between the two states in the dark-state manifold and the  intermediate state. If this is done using a so-called counter-intuitive pulse sequence, one can transport population from one state in the dark-state manifold to the other state within that manifold with perfect fidelity.
However, in our system it is not suitable to apply conventional STIRAP as the  cavity-magnon coupling, $g_{ma}$, is constant, i.e.~it cannot be modulated in a time-dependent manner. 
We therefore present in the following a modified STIRAP method 
which allows 
state transfer by modulating the detunings. 

\begin{figure}[t] 
\begin{center}
\setlength{\unitlength}{1cm}
\begin{picture}(8.5,14)
\put(-1,9){\includegraphics[width=.38\columnwidth]{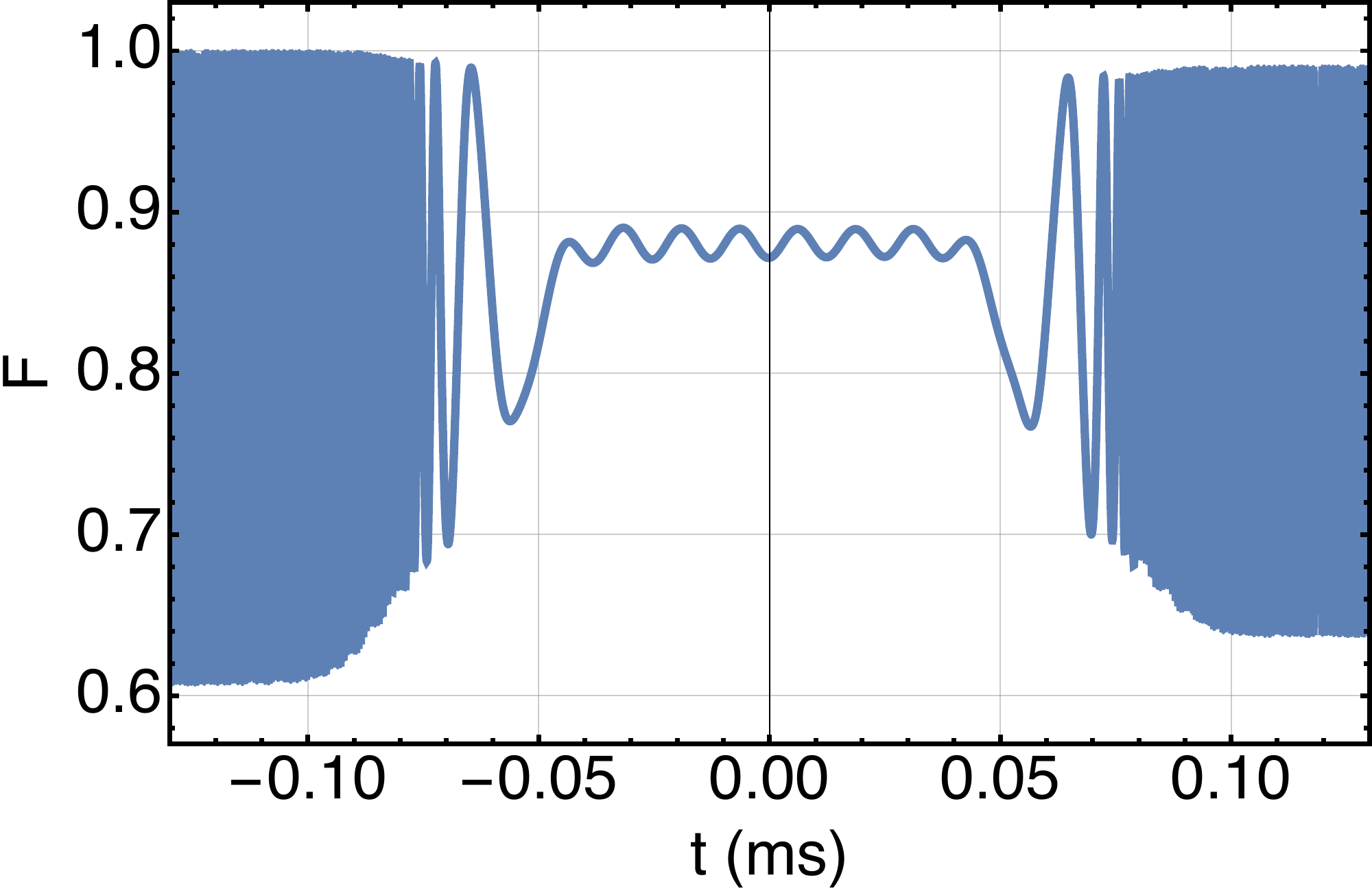}}
\put(-1.5,12.3){(a)}
\put(6,9){\includegraphics[width=.38\columnwidth]{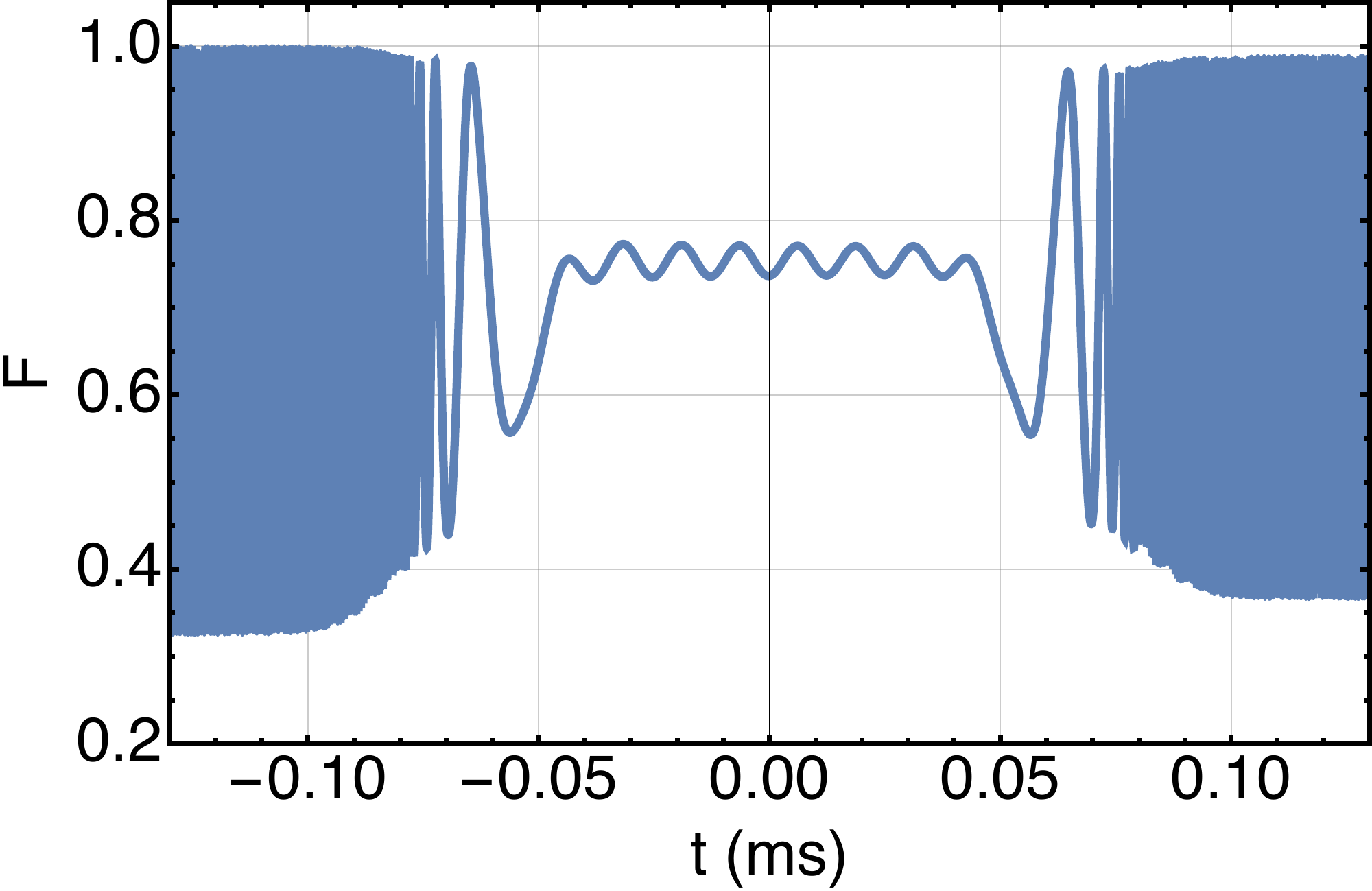}}
\put(5.5,12.3){(b)}
\put(2.5,4.55){\includegraphics[width=.38\columnwidth]{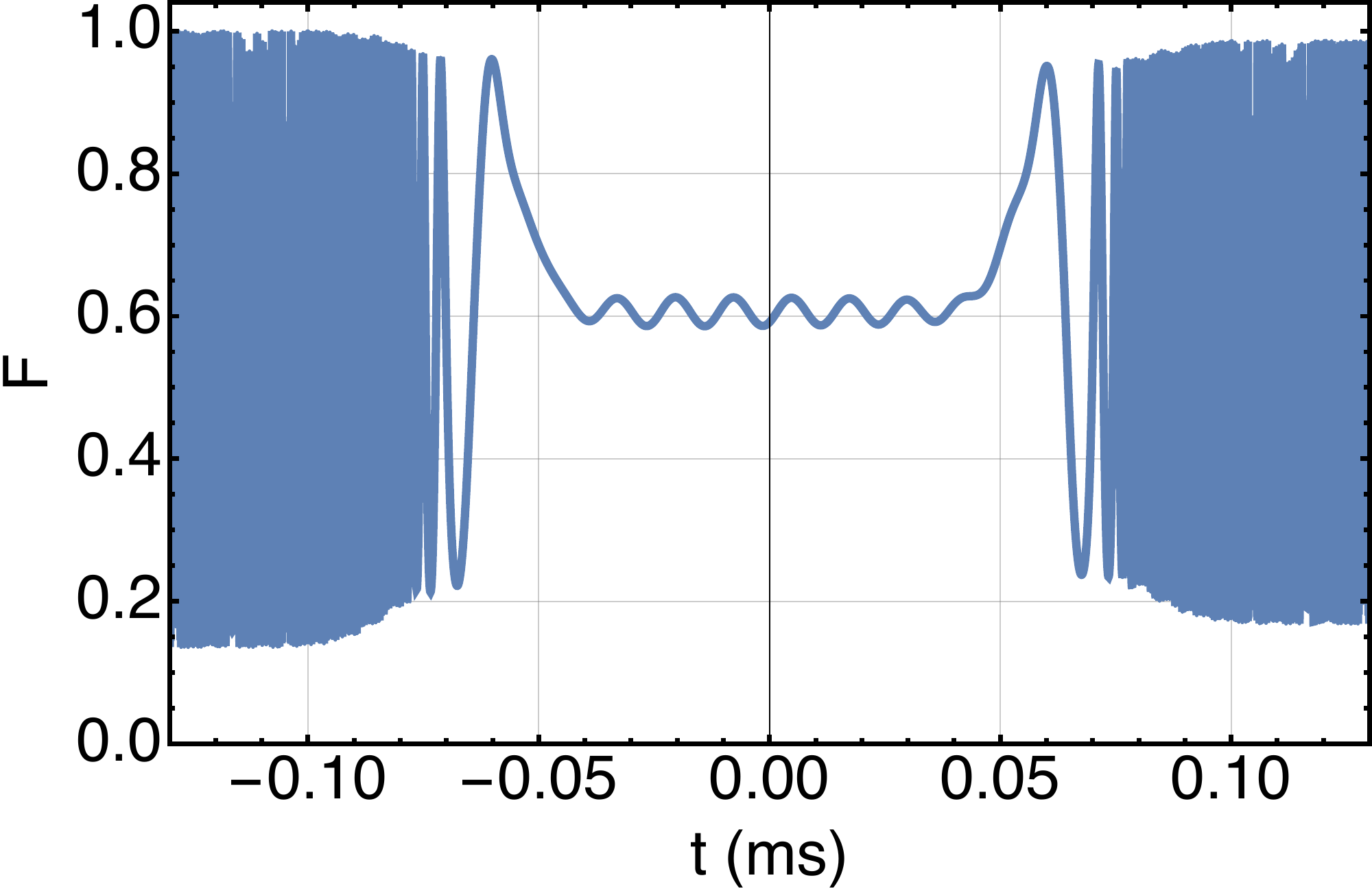}}
\put(1.9,7.8){(c)}
\put(-0.5,-2.3){\includegraphics[width=.75\columnwidth]{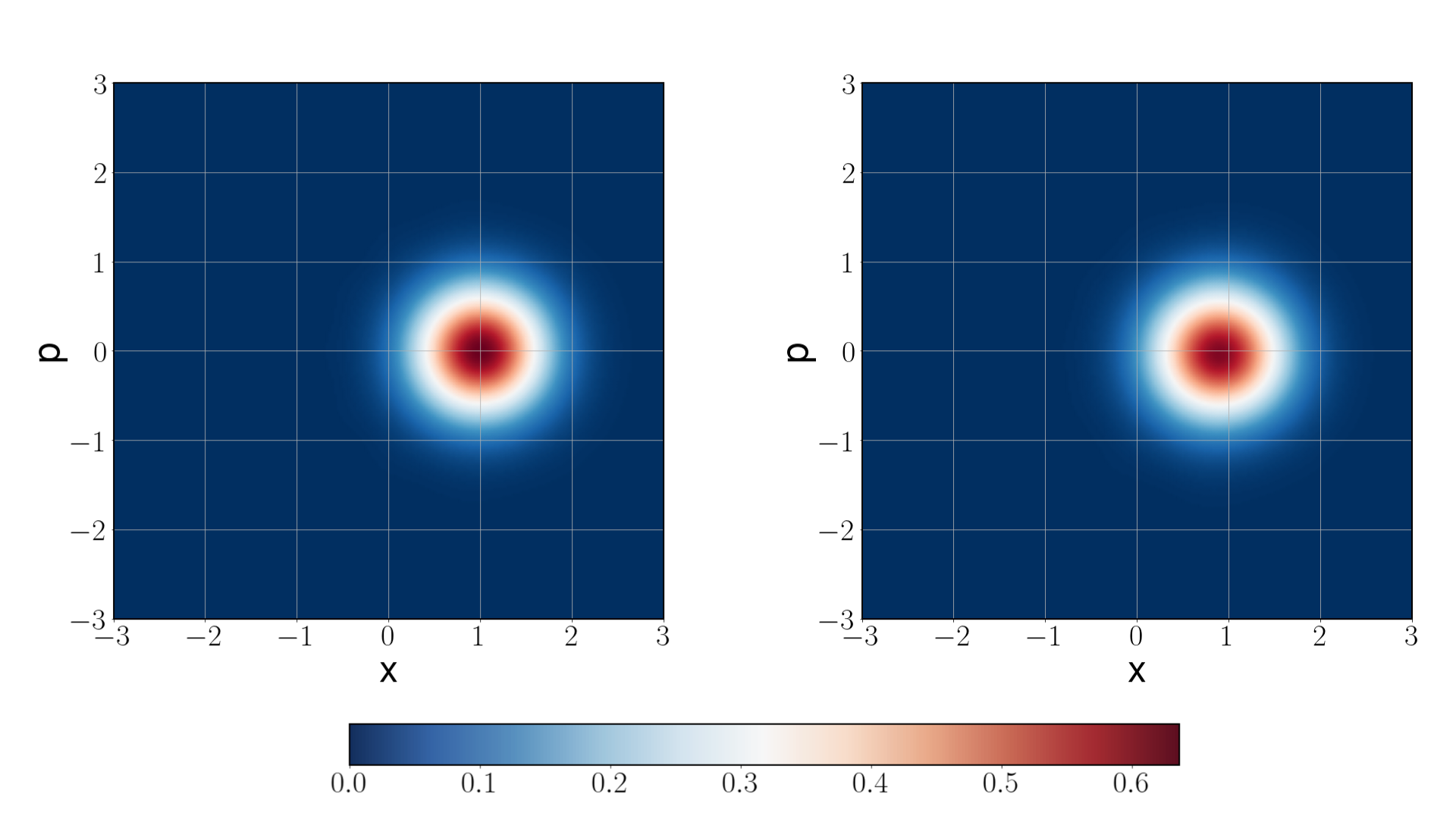}}
\put(-0.5,3.4){(d)}
\put(5.5,3.4){(e)}
\end{picture}
\vspace{1.8cm}
\end{center}
\caption{Transfer fidelities, $F(t)$, for an input coherent state in the cavity with (a) $\alpha=0.5$, (b) $\alpha=0.75$ and (c) $\alpha=1$. Panel~(d) shows the Wigner function of the input coherent state in the cavity mode with $\alpha=1$ and panel~(e) shows the Wigner function of the state in the cavity mode after the transfer. The phonon, magnon and cavity mode frequencies are considered as: $\omega_b/2\pi=10$ MHz, $\omega_a/2\pi=\omega_m/2\pi=10$ GHz.
The pulse parameters are considered as: $2\pi \Omega_0/\omega_b=0.1$, $T = 0.01$ ms, $t_{c_1} = -0.061$ ms, $t_{c_2} = 0.061$ ms, $\tau_{ch} = 0.016$ ms, $\kappa_{\delta} = 14.05$, $\tau = 0.011$ ms, $h_{\delta} = 13.94$. 
The bath temperature is considered as $T_{\rm th}=1$ mK, and the damping rates are considered as: $\kappa_b=100$ Hz and $\kappa_m=10$ kHz.
}
\label{fig:coherent_transfer}
\end{figure}
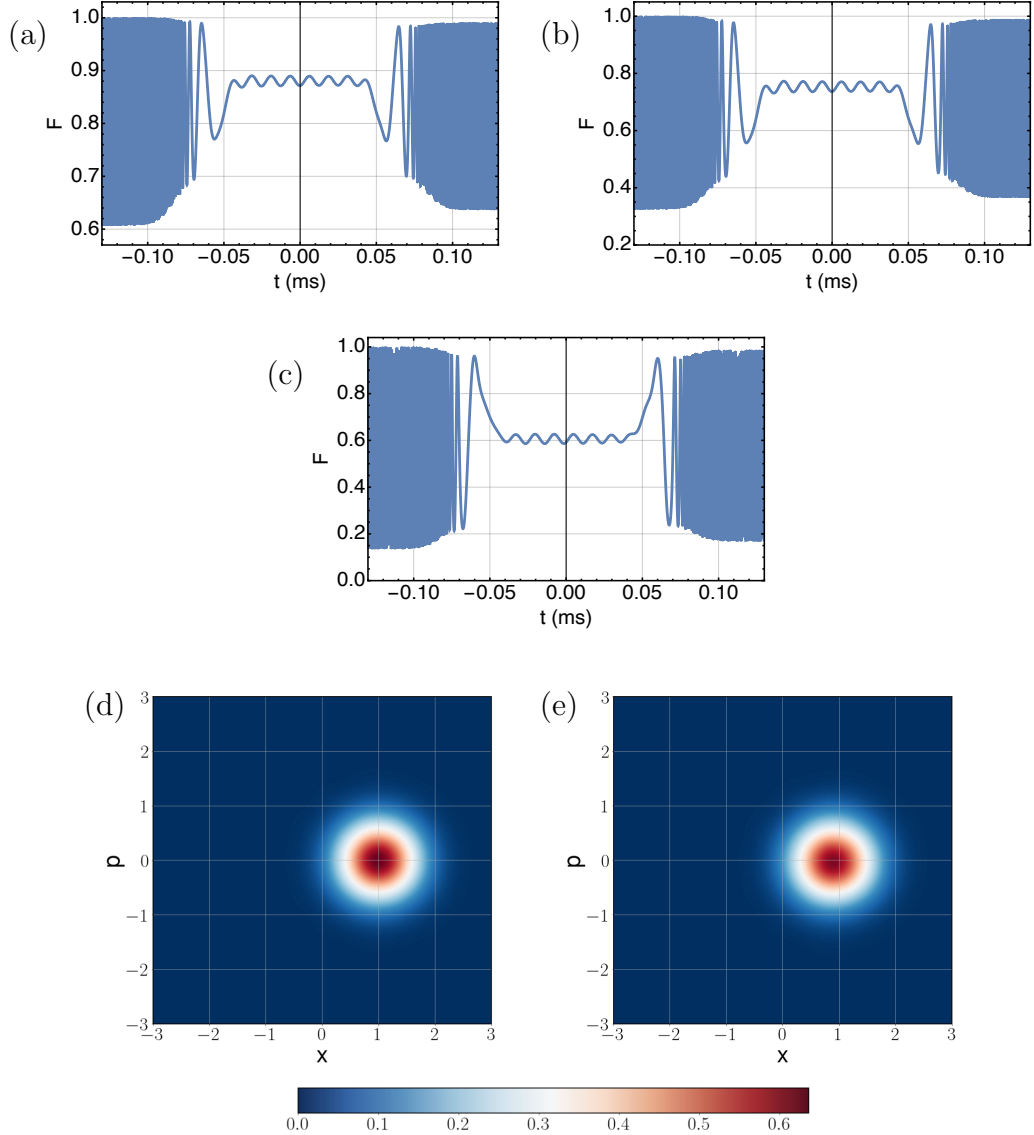

\begin{figure}[t] 
\begin{center}
\setlength{\unitlength}{1cm}
\begin{picture}(8.5,14)
\put(-1,9){\includegraphics[width=.38\columnwidth]{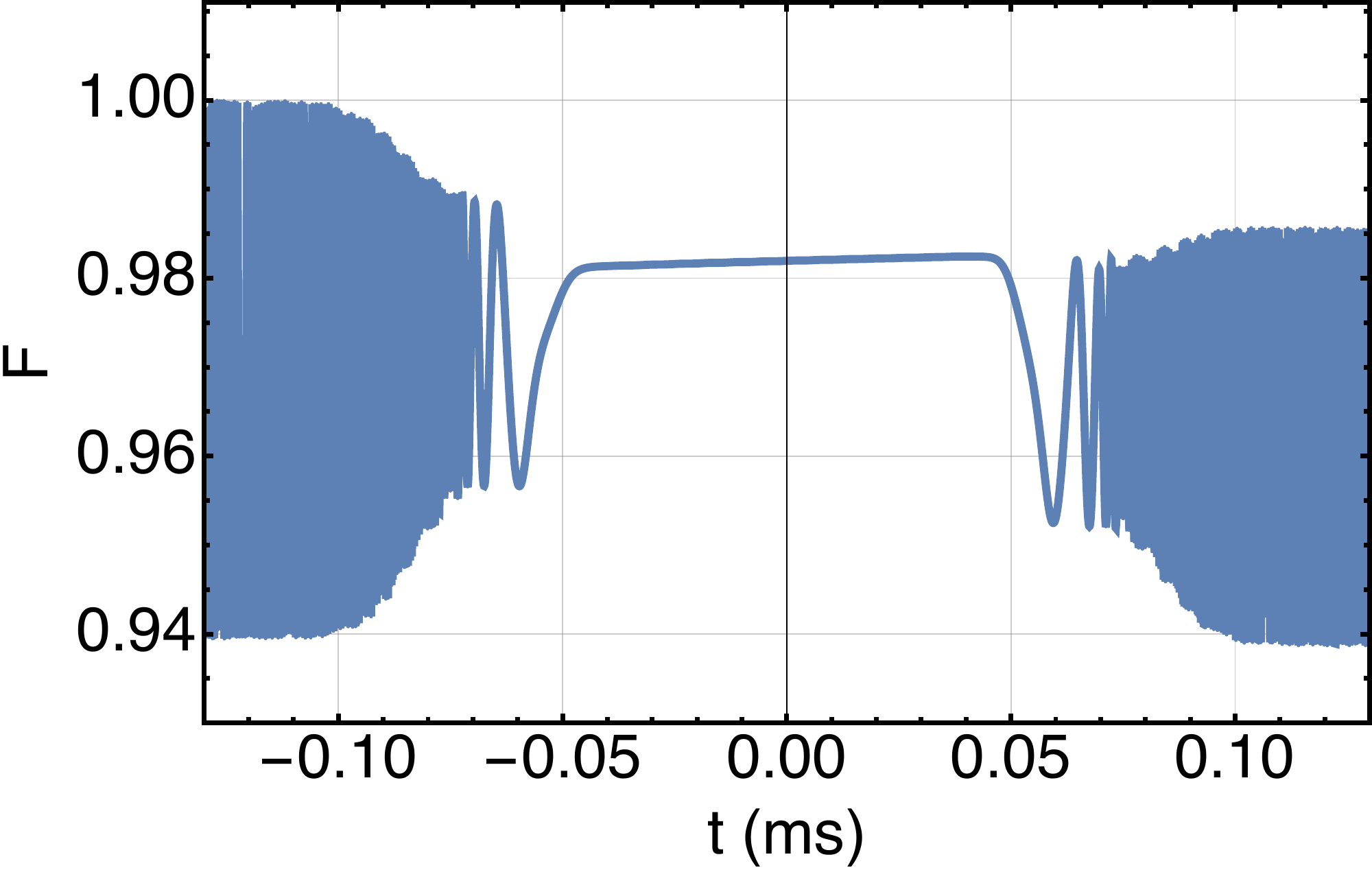}}
\put(-1.5,12.3){(a)}
\put(6,9){\includegraphics[width=.38\columnwidth]{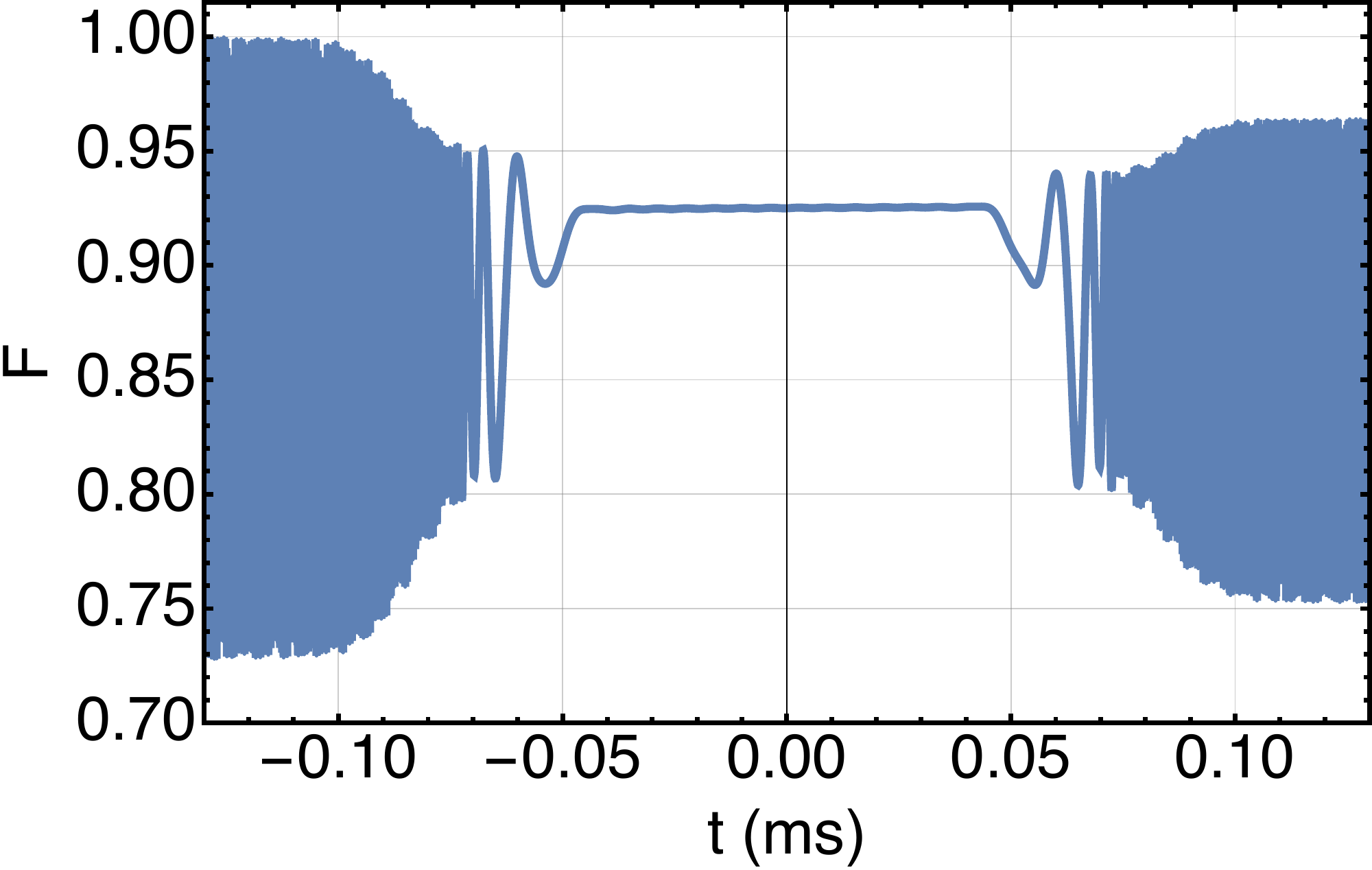}}
\put(5.5,12.3){(b)}
\put(2.5,4.55){\includegraphics[width=.38\columnwidth]{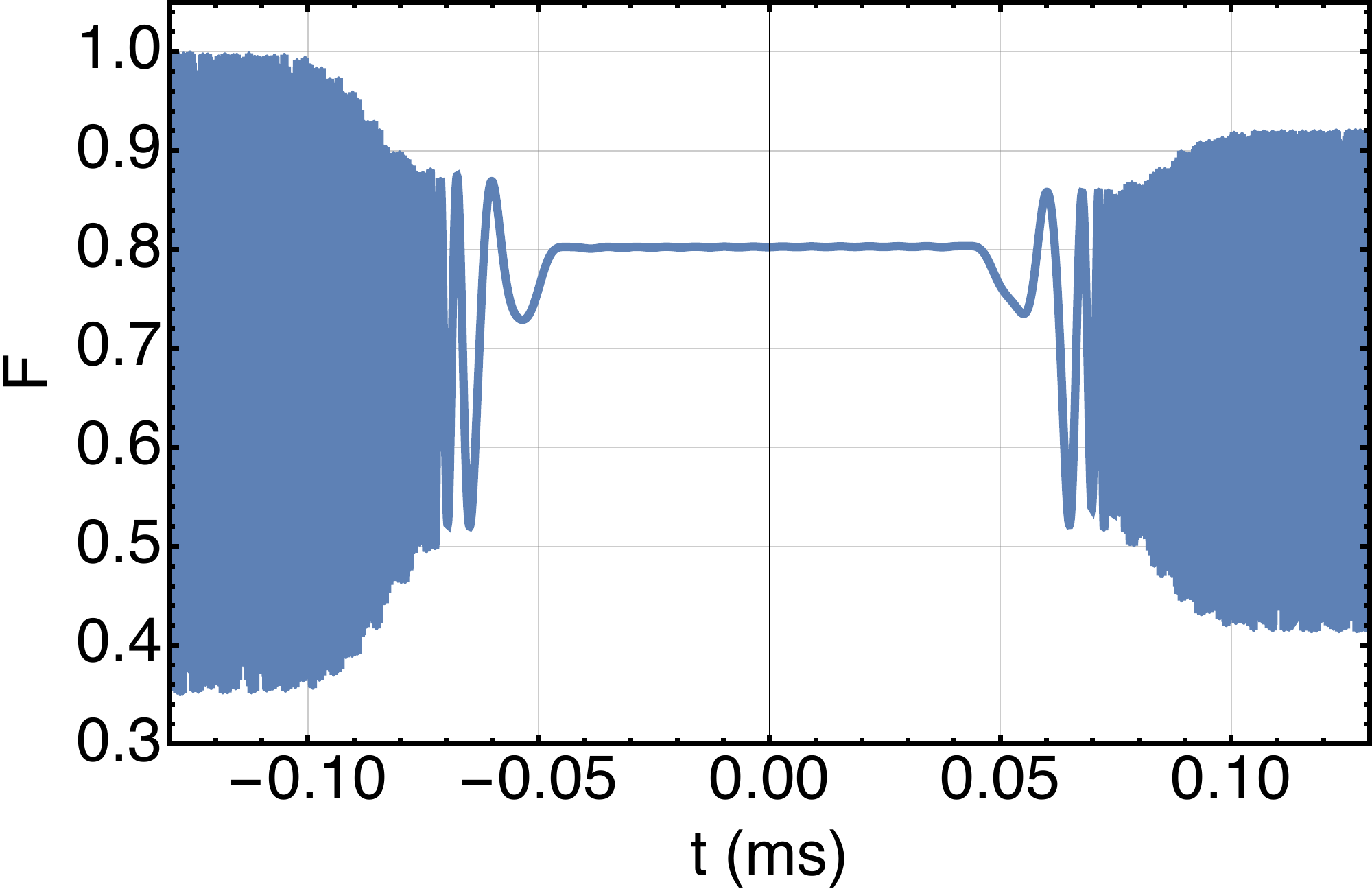}}
\put(1.9,7.8){(c)}
\put(-0.5,-2.3){\includegraphics[width=.75\columnwidth]{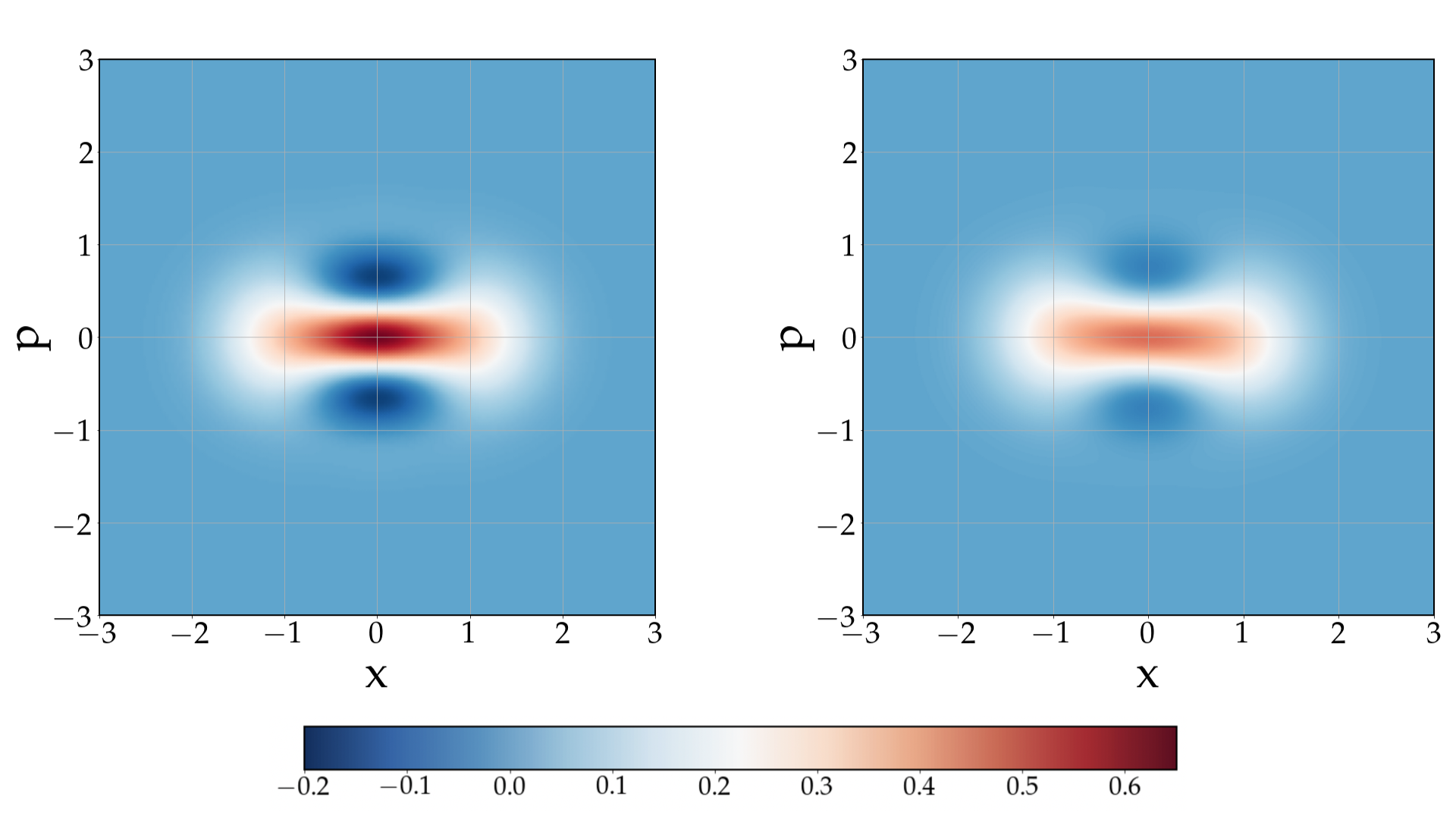}}
\put(-0.5,3.4){(d)}
\put(5.5,3.4){(e)}
\end{picture}
\end{center}
\vspace{1.8cm}
\caption{Transfer fidelities for an input cat state in the cavity mode with (a) $\alpha=0.5$, (b) $\alpha=0.75$, and (c) $\alpha=1$. Figures (d) and (e) show the input and output Wigner functions for the cavity mode with input cat state with $\alpha=1$.
Parameters are same as in Fig.~\ref{fig:coherent_transfer}. 
}
\label{fig:cat_transfer}
\end{figure}

In this scheme the cavity-magnon coupling $g_{ma}\equiv\Omega_s/2=\Omega_0/2$, which is called the `Stokes' coupling in the context of atomic STIRAP, is constant in time and the magnomechanical coupling $G_{mb}(t) \equiv \Omega_p(t)/2$, which is generally known as the `Pump' coupling in STIRAP, is modulated as
\begin{equation}
\label{eq:pulse} 
\Omega_p(t)= \Omega_{p_1}(t) + \Omega_{p_2}(t).
\end{equation}
Here
\begin{equation}
\label{eq:pulse1a}
\Omega_{p_1}(t)= \Omega_0 \mathrm{e}^{-\left(\frac{t-t_{c_1}}{T}\right)^2},
\end{equation}
and 
\begin{equation}
\label{eq:pulse2}
\Omega_{p_2}(t)= \Omega_0 \mathrm{e}^{-\left(\frac{t-t_{c_2}}{T}\right)^2},
\end{equation}
are Gaussian-shaped pulses centered at the times $t_{c_1}$ and $t_{c_2}$, with width $T$, and amplitude $\Omega_0$. Application of the pulse $\Omega_{p_1}(t)$ transfers the state from the cavity to the mechanical mode, whereas the application of the pulse $\Omega_{p_2}(t)$ brings the state back to the cavity.
The cavity and magnon detunings are considered as
\begin{eqnarray}
\label{eq:detunings}
\nonumber
\delta_m(t) &=- \kappa_{\delta} h_{\delta} \frac{\Omega_0}{2} \left[\tanh\left(\frac{t-\tau}{\tau_{ch}}\right) + \tanh\left(\frac{t+\tau}{\tau_{ch}}\right)\right]\; ,\\
\delta_a(t) &= -(\kappa_{\delta} -1)h_{\delta} \frac{\Omega_0}{2} \left[\tanh\left(\frac{t-\tau}{\tau_{ch}}\right) + \tanh\left(\frac{t+\tau}{\tau_{ch}}\right)\right]\;,
\end{eqnarray}
so that
\begin{eqnarray}
\delta_m(t)- \delta_a(t)&=\delta_s(t) 
&=-h_{\delta} \frac{\Omega_0}{2} \left[\tanh\left(\frac{t-\tau}{\tau_{ch}}\right) + \tanh\left(\frac{t+\tau}{\tau_{ch}}\right)\right]\;. 
\end{eqnarray}
The shapes of these coupling and detuning pulses are shown in Fig.~\ref{fig:schematic_model}(b). 
The choice of these pulse shapes for the magnomechanical coupling and the detunings can be understood by looking at the instantaneous eigenvalues of the system.
In the  rotating wave 
approximation the Hamiltonian  (\ref{RWA1})
is given by 
\begin{eqnarray}
H=
\left[ {\begin{array}{ccc}
	0 								& 	\Omega_p (t)/2							& 0	\\
	\Omega_p^* (t)/2								& 	\delta_m (t)					& \Omega_s/2	\\
	0		& 	\Omega_s/2 	& \delta_a(t) 						\\
	\end{array} } \right]\;\;
\end{eqnarray}
which is similar to the Hamiltonian used in STIRAP in a three-level atom with states $|0,1,2\rangle$, where the transfer is sought from $|0\rangle$ to $|2\rangle$ without populating $|1\rangle$. In our system this corresponds to the transfer from $|a\rangle$ to $|b\rangle$ and vice versa in the retrieval stage of the scheme, with no occupation of $|m\rangle$ during the storage stage. The instantaneous eigenvalues ($\lambda_0$, $\lambda_1$,$\lambda_2$) of the Hamiltonian in Eq.~(11), when the time modulated pulses are applied, are shown in Fig.~\ref{fig:lossless_transfer}(a) (solid lines). If we consider the magnomechanical coupling $G_{mb}(t)$ to vanish (i.e.~$\Omega_p(t)=0$), the Hamiltonian is given by
\begin{eqnarray}
H_s=
\left[ {\begin{array}{ccc}
	0 								& 	0							& 0	\\
	0								& 	\delta_m (t)					&  \Omega_0/2	\\
	0		& 	 \Omega_0^*/2	& \delta_a(t) 						\\
	\end{array} } \right]\;\;,
\end{eqnarray}
and it acts only on the 
cavity-magnon subspace, i.e.~it does not involve the mechanical mode. This yields the asymptotic eigenstates $\ket{s_0(t=\pm \infty)}$, and $\ket{s_\pm(t=\pm \infty)}$, where
\begin{eqnarray}
\ket{s_+(-\infty)} &\simeq \ket{\Psi_b}\to\ket{s_+(+\infty)} \simeq \ket{\Psi_a}\;\;,
\\
\ket{s_-(-\infty)} &\simeq \ket{\Psi_a}\to\ket{s_-(+\infty)} \simeq \ket{\Psi_b}\;\;.
\end{eqnarray}
Here $\ket{\Psi_{a}}$ ($\ket{\Psi_{b}}$) are the states of the microwave cavity (mechanical mode) and  the corresponding eigenvalues are
\begin{equation}
\label{eq:diabatic-eigenvectors}
S_0 = 0, \quad
S_{\pm} = \frac{\delta_a+\delta_m}{2} \pm \frac{\sqrt{(\delta_a -\delta_m) ^2 + \Omega_0^2}}{2}\;\;.
\end{equation}
The time evolution of the eigenvalues of this Stokes Hamiltonian results in the eigenvalues $S_\pm (t)$ crossing the eigenvalue $S_0$ twice at $t\sim t_{c_1}$  and $t\sim t_{c_2}$ as shown by the dashed lines in Fig.~\ref{fig:lossless_transfer}(a). However, the application of the magnomechanical coupling $\Omega_p$ lifts the degeneracy between the eigenstates $S_-$ and $S_0$ in the first avoided crossing, and $S_0$ and $S_+$, in the second avoided crossing (shown by the solid lines in Fig.~\ref{fig:lossless_transfer}(a)), which leads to state transfers at these two points. The corresponding population transfers at these two points for an initial Fock state in the cavity are depicted in Fig.~\ref{fig:lossless_transfer}(b), where we show the time evolution of the phonon, magnon and photon mode occupancy, $N_b$, $N_m$ and $N_a$ considering an initial state $(N_a, N_m, N_b)=(1,0,0)$. This behavior is obtained by solving the Schr\"odinger equation without considering any coupling of the system to external baths. 
One can see that the population is transferred with nearly 100\% fidelity from the cavity mode $a$ to the mechanical mode $b$ and again back to the mode $a$. The population in the magnon mode, $m$, is briefly non-zero around $t\sim t_{c_1}$ and $t\sim t_{c_2}$, however quickly returns to vanishing occupancy, leading to a complete transfer between the cavity and mechanics, despite a vast difference in frequencies between them.

{\color{black}
Next, we consider the situation where the cavity mode is initially in a coherent state $|\Psi_i\rangle={\cal D}(\alpha)|0\rangle$, where ${\cal D}(\alpha) = e^{\alpha a^\dagger -\alpha ^* a}$.
 The performance of our quantum memory scheme is evaluated in terms of the 
fidelity, 
$F=\sqrt{\langle \Psi_i |\rho_a(t)|\Psi_i \rangle}$}, 
which measures the overlap of the density matrix of the instantaneous cavity state, $\rho_a(t)={\rm Tr_{m,b}}[\rho(t)]$, with the initial cavity state $|\Psi_i\rangle$. For an initial coherent state with $\alpha=1$ the transfer and retrieval processes are shown in Fig.~\ref{fig:lossless_transfer}(c) and values of $F\sim 1$ indicate that the transferred state closely resembles the original cavity state intended to be transferred, stored and retrieved \cite{nielsen2011quantum}. In Fig.~\ref{fig:lossless_transfer}(c) the rapid oscillations in the fidelity at the starting and ending portions of the pulse are due to the rapid rotation of the cavity state in phase space with large detunings. By halting the pulse at the opportune time we obtain near-unit fidelity, which shows near-perfect transfer and retrieval.

It is to be noted that until now in our analysis, we have not taken into account any damping existing in the system. We will therefore, in the following, study the state transfer dynamics for various input states in a realistic open system by coupling all modes to a thermal bath. 
\begin{figure}[t] 
\begin{center}
\setlength{\unitlength}{1cm}
\begin{picture}(8.5,14)
\put(-1,9){\includegraphics[width=.4\columnwidth]{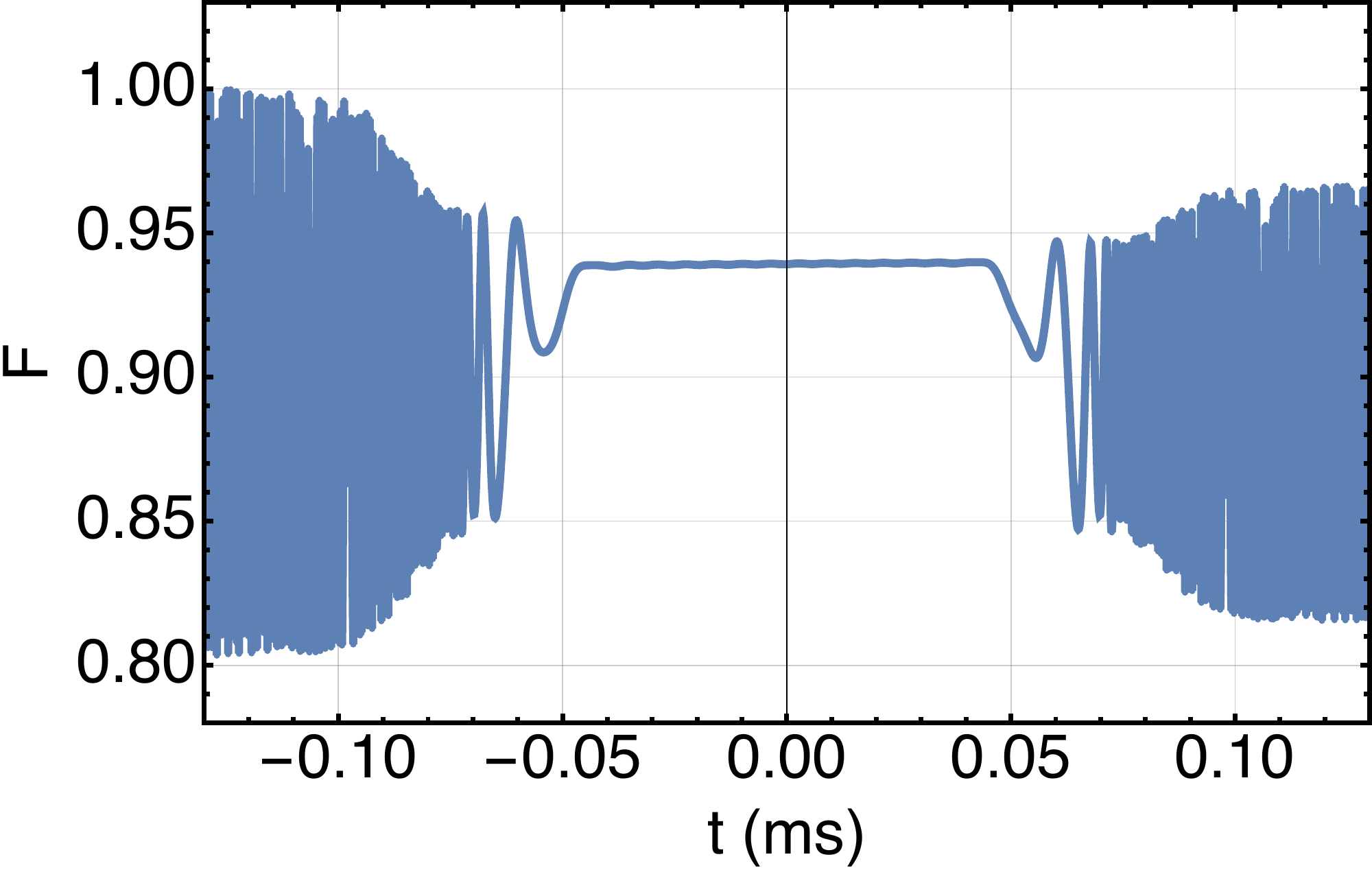}}
\put(-1.5,12.3){(a)}
\put(6,9){\includegraphics[width=.4\columnwidth]{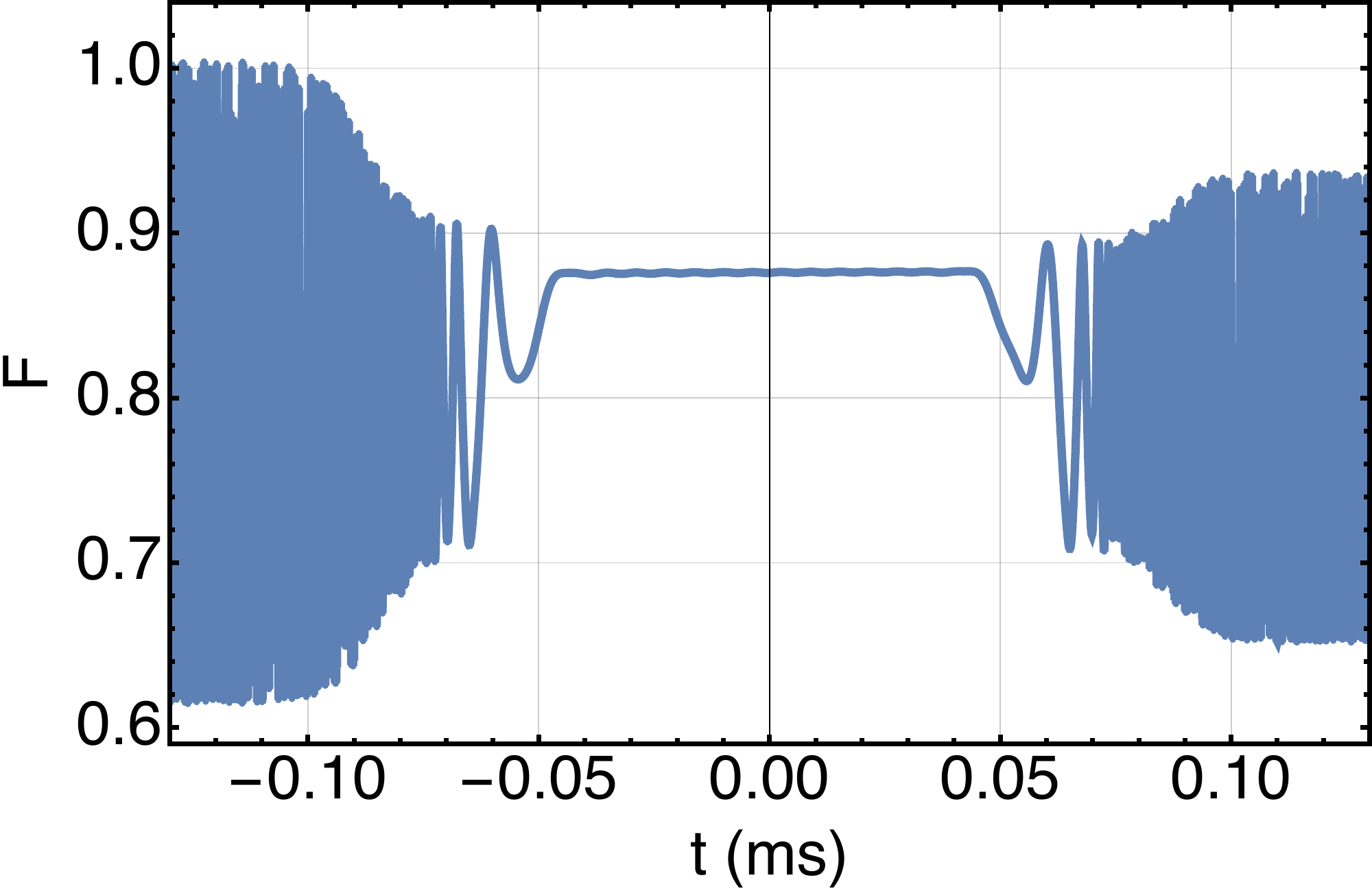}}
\put(5.5,12.3){(b)}
\put(2.5,4.55){\includegraphics[width=.4\columnwidth]{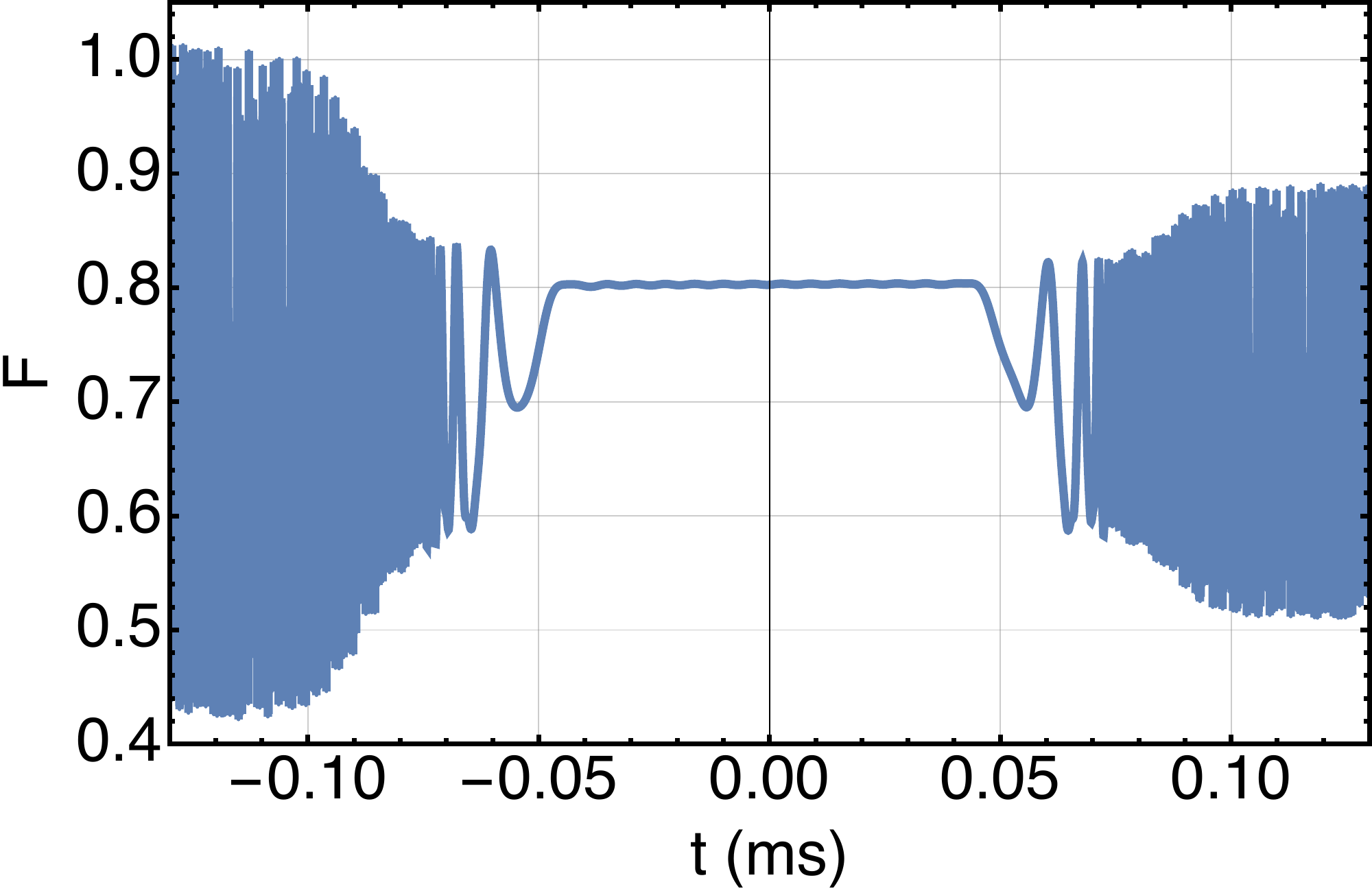}}
\put(1.9,7.8){(c)}
\put(-0.5,-2.3){\includegraphics[width=.75\columnwidth]{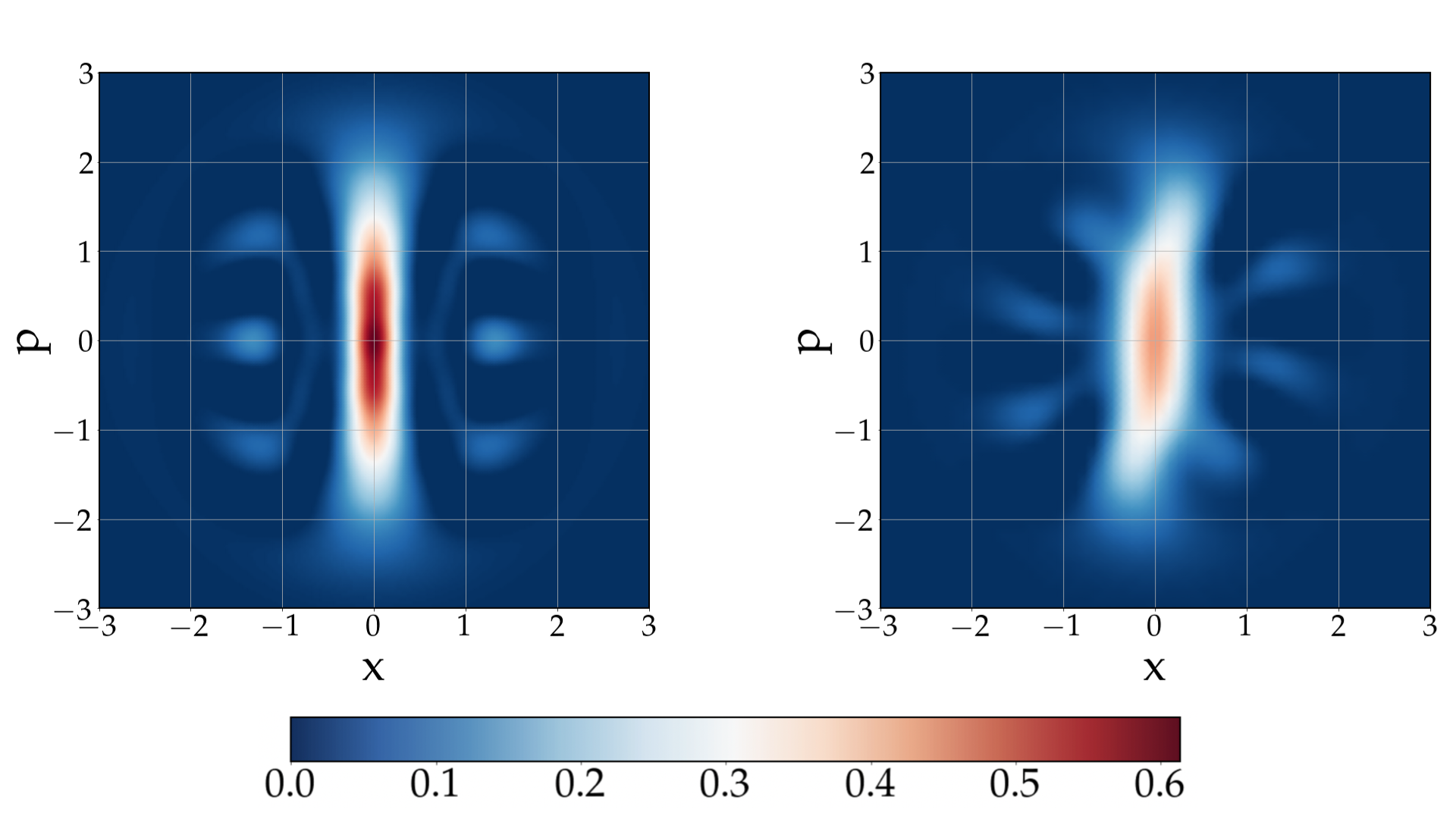}}
\put(-0.5,3.3){(d)}
\put(5.5,3.3){(e)}
\end{picture}
\end{center}
\vspace{1.8cm}
\caption{The transfer fidelities for an input squeezed vacuum state in the cavity mode with (a) $r=0.5$, (b) $r=0.75$, and (c) $r=1$. Figures (d) and (e) show the Wigner functions for the initial and retrieved  cavity mode with input squeezed vacuum state with $r=1$.
Parameters are same as in Fig.~\ref{fig:coherent_transfer}. 
}
\label{fig:squeezed_transfer}
\end{figure}
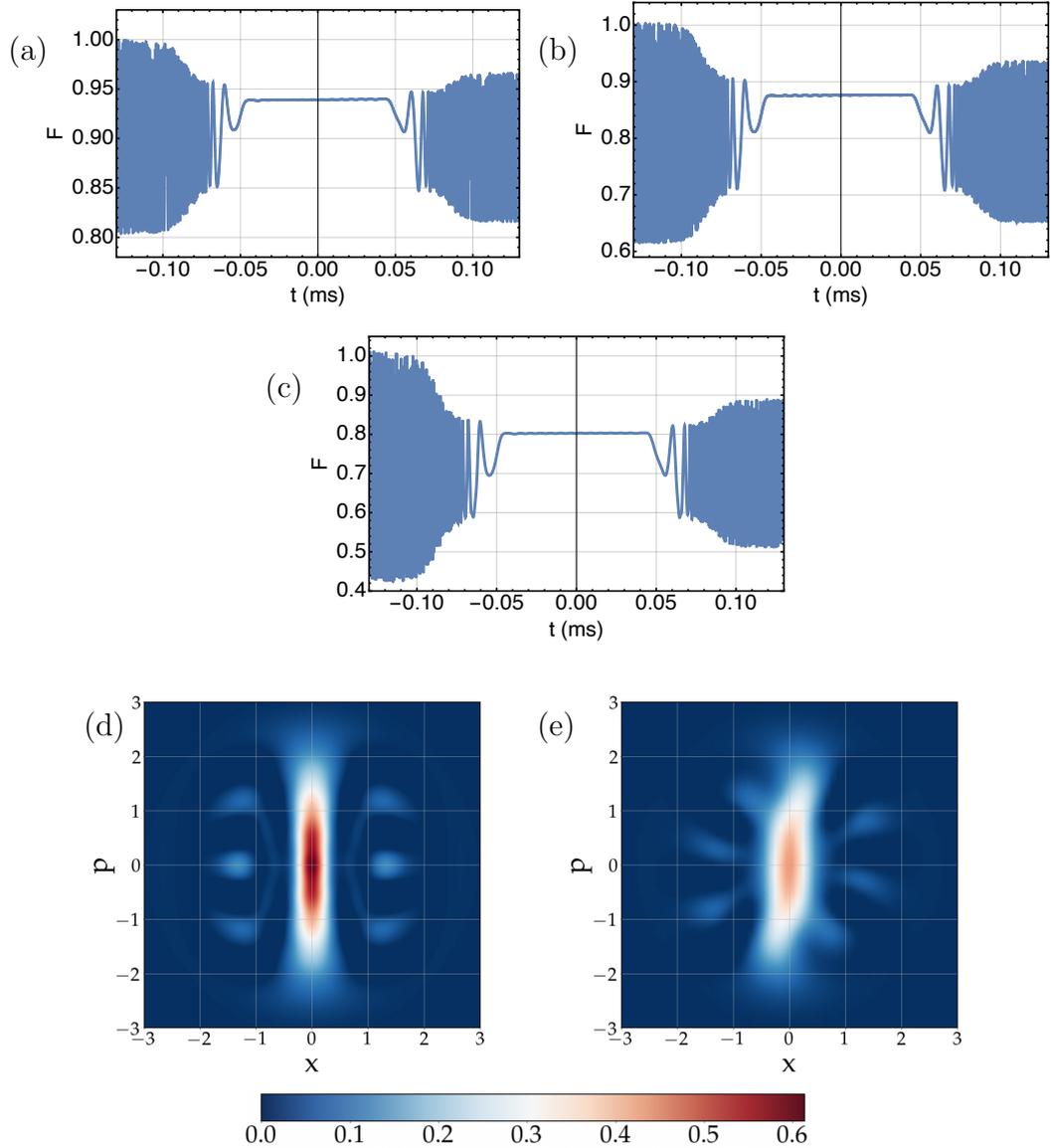
We consider the open quantum system dynamics using the quantum master equation,
\begin{eqnarray}						
\dot{\rho}= & i\left[\rho, H\right]+
\left\{\kappa_a\left(\bar{n}_{a} +1\right){\mathcal{L}}[a_1]+\kappa_a \bar{n}_{a}{\mathcal{L}}[a_1^\dagger]\right.\nonumber\\
&+\kappa_m\left(\bar{n}_{m} +1\right){\mathcal{L}}[m_1]+\kappa_m \bar{n}_{m}{\mathcal{L}}[m_1^\dagger]\nonumber\\
&\left.+\kappa_b\left(\bar{n}_{b} +1\right){\mathcal{L}}[b_1]+\kappa_b \bar{n}_{b}{\mathcal{L}}[b_1^\dagger]\right\}\rho, \label{master_eq}
\end{eqnarray}%
with ${\mathcal{L}}[A]\rho\equiv A\rho A^\dagger-1/2\,\{A^\dagger A,\rho\}$ representing the dissipation and noises in the system. 
We consider the phonon, magnon and cavity mode frequencies as $\omega_b/2\pi=10$ MHz and $\omega_a/2\pi=\omega_m/2\pi=10$ GHz, and the bath temperature is chosen to be $T_{\rm th}=1$ mK. 
Since the cavity and magnon modes oscillate at high frequencies, coupling these to a thermal bath at mK temperatures yields almost zero thermal occupancy, however the phonon mode is occupied. The damping rates are considered as $\kappa_b=100$ Hz and $\kappa_m=10$ kHz. In the following we also consider $\kappa_a=0$, as we are only interested in the loss of fidelity caused by the transfer and storage processes. We start the transfer from an initial state given by $|\alpha, 0, 0\rangle$, where the cavity mode contains a coherent state, the magnon mode occupation is zero and the mechanical mode has been precooled to the ground state \cite{teufel2011sideband,schliesser2006radiation,sarma2020optomechanical}.    

In Fig.~\ref{fig:coherent_transfer}(a-c) we show the instantaneous fidelities for the transfer and retrieval processes for the coherent state from the cavity to the mechanical mode and back to the cavity mode for various values of $\alpha$. 
For all examples the retrieval fidelity is nearly $100 \%$.
To further quantify the
effectiveness of the photon-phonon-photon transfer, we 
calculate the Wigner function for the initial cavity state and for the cavity state after the transfer. One can see from Figs.~\ref{fig:coherent_transfer}(d) and (e)
that both these Wigner functions are virtually identical
with only a small shift after storage and retrieval towards the vacuum state.

Next, we consider two slightly more complicated states. In Fig.~\ref{fig:cat_transfer}, we show the storage and retrival fidelity  of an initial cat state given by $|\Psi_i\rangle = {\cal N}\, (|\alpha \rangle + |-\alpha \rangle )$, where ${\cal N}$ is a normalization parameter. Panels (a)-(c) show the results for the transfer fideltiy for $\alpha=0.5, 0.75, 1$  and panels (d) and (e) show the Wigner functions for the cavity state before and after the transfer 
for the state with $\alpha=1$. We also consider in Fig.~\ref{fig:squeezed_transfer} the squeezed vacuum state as the initial state in the cavity given by $|\Psi_i\rangle = e^{(r a^{\dagger^2}-r^* a^2)/2}|0\rangle$, where $r$ is the squeezing parameter. The storage fidelities for three different squeezing strengths $r=0.5, 0.75, 1$ are shown in Figs.~\ref{fig:squeezed_transfer}(a)-(c) and panels (d) and (e) depict the Wigner functions for the cavity state before and after the transfer with $r=1$.
One can see in both cases that, even though the retrieval fidelity is rather high, it does no longer reach the value of $F=1$. For the cat state it decreases with increasing values of $\alpha$, which suggests that the overall fluctuations of the two terms, 
$|\alpha\rangle$ and $|-\alpha\rangle$ become more prominent and are the reason for the loss of fidelity. Similarly for squeezed states the decoherence distorts the squeezed feature in phase space, resulting in the lower fidelity. As $r$ increases the distortion is more pronounced, which is visible in the shape of the Wigner function of the retrieved photon in Fig.~\ref{fig:squeezed_transfer}(e). However, for all the input states described here, the storage fidelity is still very high which shows that the transfer is highly efficient.

 \begin{figure}[t] 
 	\begin{center}
 		\setlength{\unitlength}{1cm}
 		\begin{picture}(8.5,14)
 		\put(-3.5,9){\includegraphics[width=.45\columnwidth]{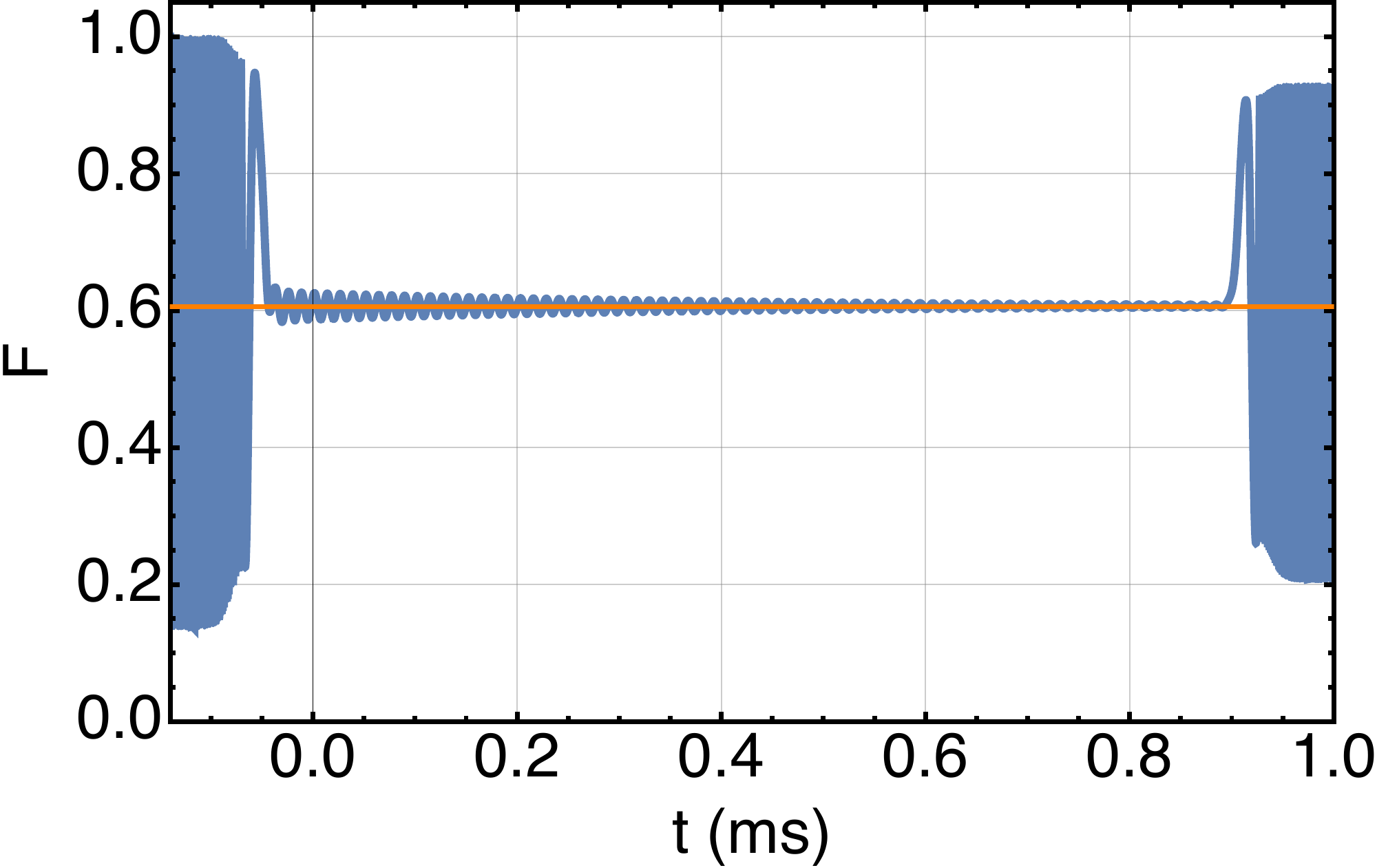}}
 		\put(-4.,12.7){(a)}
 		\put(4.7,9){\includegraphics[width=.52\columnwidth]{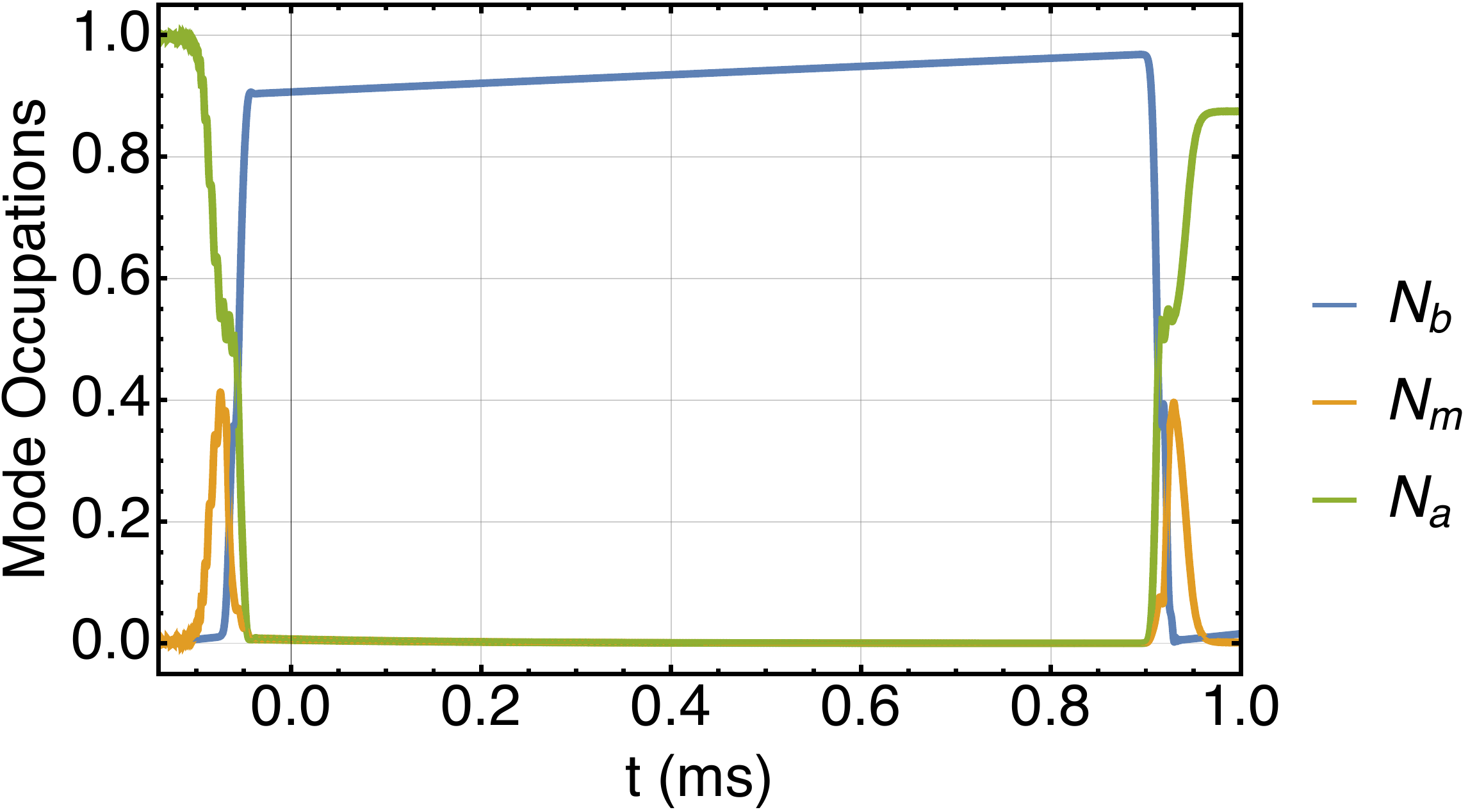}}
 		\put(4.,12.7){(b)}
 		\put(3,4.55){\includegraphics[width=.4\columnwidth]{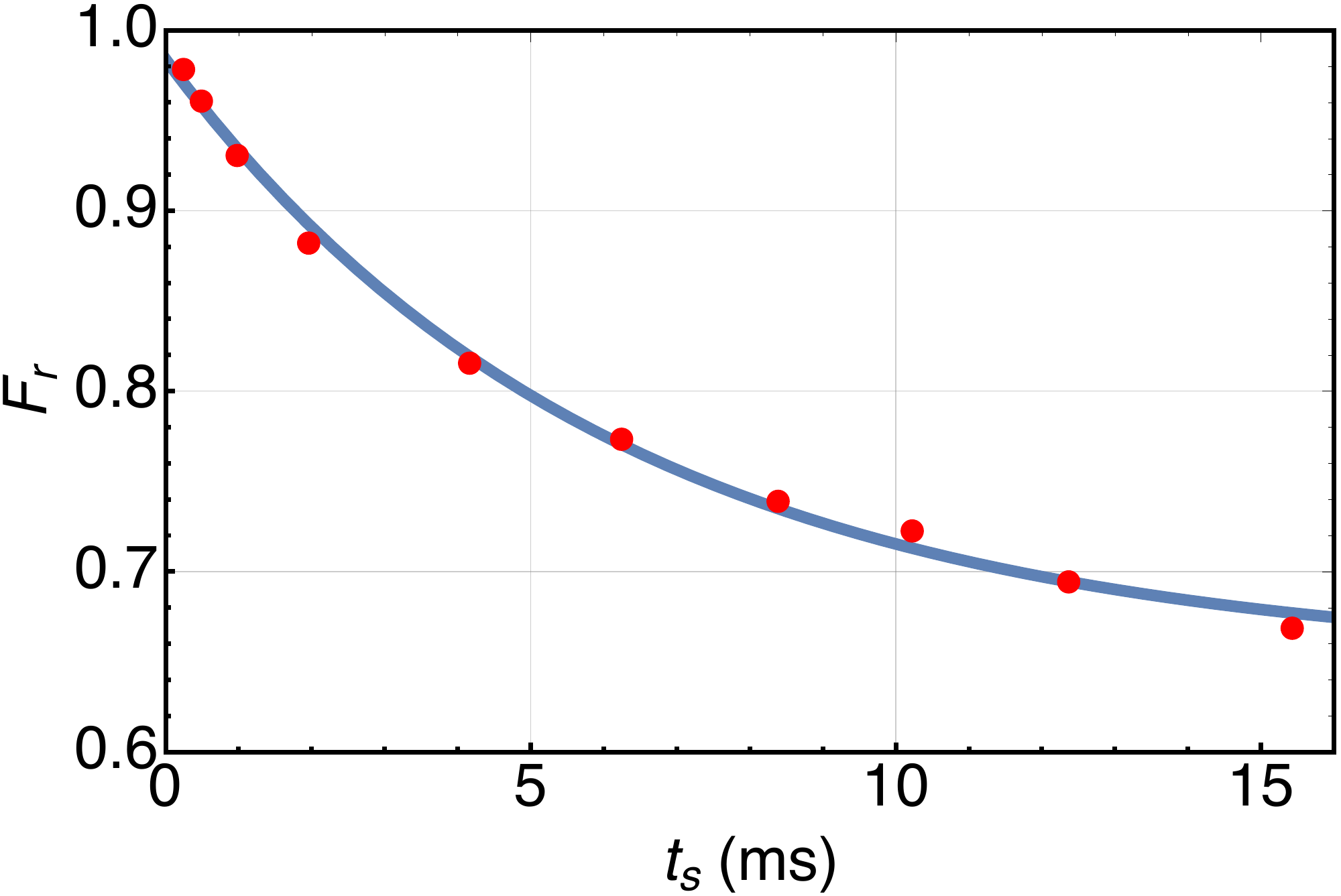}}
 		\put(2.5,7.9){(c)}
		\end{picture}
 	\end{center}
 	\vspace{-5cm}
 	\caption{(a) Transfer fidelity with delay time $\Delta t = 14\times t_{c_2}$ for an input coherent state in the cavity mode with $\alpha=1$. The phonon, magnon and cavity mode frequencies are considered as: $\omega_b/2\pi=10$ MHz, $\omega_a/2\pi=\omega_m/2\pi=10$ GHz.
The pulse parameters are considered as: $2\pi \Omega_0/\omega_b=0.1$, $T = 0.01$ ms, $t_{c_1} = -0.061$ ms, $t_{c_2} = 0.061$ ms, $\tau_{ch} = 0.016$ ms, $\kappa_{\delta} = 14.05$, $\tau = 0.011$ ms, $h_{\delta} = 13.94$.
The bath temperature $T_{\rm th}=1$ mK, and the damping rates are taken as: $\kappa_b=100$ Hz and $\kappa_m=10$ kHz.  The solid orange line shows the vacuum occupation which is reached by the cavity mode after the state transfer to the mechanical mode as the storage time is chosen to be very long. (b) The corresponding variation of the mode occupations with respect to the added delay time, $\Delta t = 14\times t_{c_2}$. (c) Variation of the retrieved maximum fidelity, $F_r$ (red dots) with respect to the storage time $t_s$, obtained with $\Delta t = (2\times t_{c_2}, 6\times t_{c_2}, 14\times t_{c_2}, 30\times t_{c_2}, 66\times t_{c_2}, 100\times t_{c_2}, 135\times t_{c_2}, 165\times t_{c_2}, 200\times t_{c_2}, 250\times t_{c_2})$. The blue line shows an exponential decay fit of the form $F_r = A e^{-t_s/t_{\rm half}} + A_0$, where $t_{\rm half}$ is the half lifetime. 
}
\label{fig:delayed_transfer}
 \end{figure}
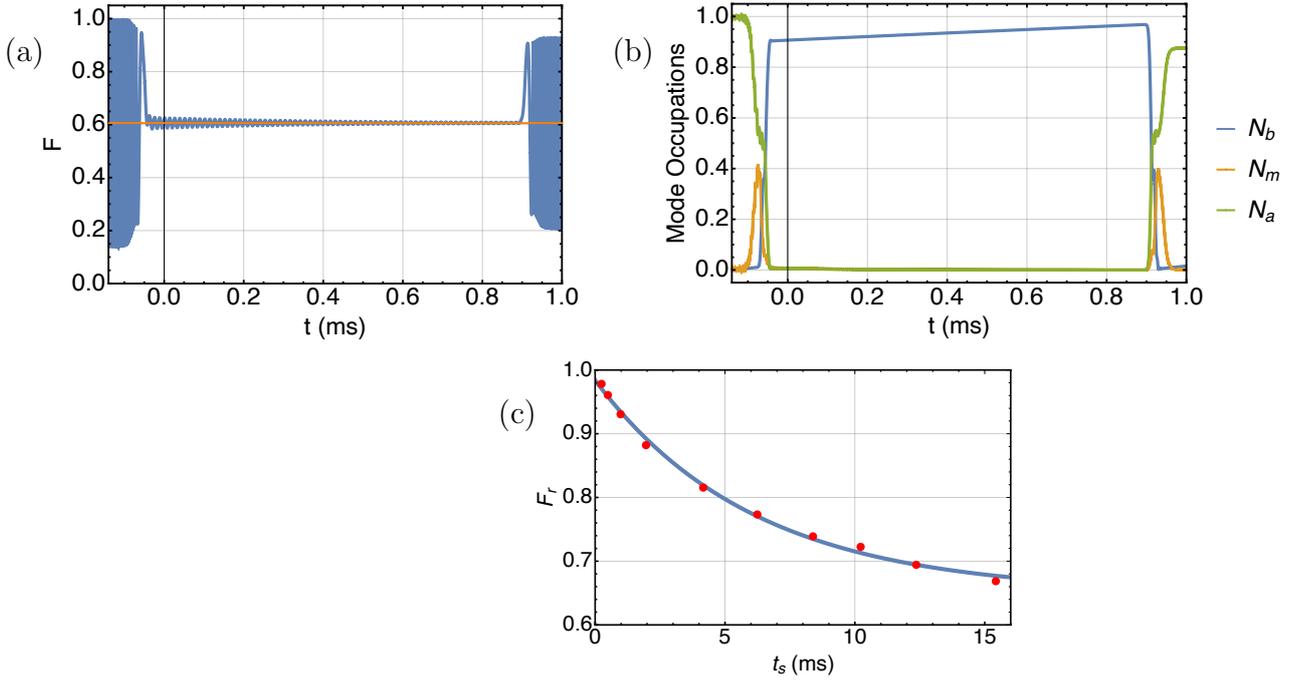

The above analysis shows that the mechanical mode can be used as a storage mode for the photonic state. However, as the fidelity inevitably decays for longer storage times, it is interesting to study how this decay happens. 
For this we introduce a time-delay of the retrieval pulse by $\Delta t$, so that
\begin{equation}
\label{eq:pulse2b}
\Omega_{p_2}(t)= \Omega_0 \mathrm{e}^{-\left(\frac{t-(t_{c_2}+\Delta t)}{T}\right)^2}.
\end{equation}
and similar for the cavity and magnon detunings 
\begin{eqnarray}
\label{eq:detunings2}
\nonumber
\delta_m(t) &=- \kappa_{\delta} h_{\delta} \frac{\Omega_0}{2} \left[\tanh\left(\frac{t-(\tau +\Delta t )}{\tau_{ch}}\right) + \tanh\left(\frac{t+\tau}{\tau_{ch}}\right)\right]\; ,\\
\delta_a(t) &= -(\kappa_{\delta} -1)h_{\delta} \frac{\Omega_0}{2} \left[\tanh\left(\frac{t-(\tau +\Delta t)}{\tau_{ch}}\right) + \tanh\left(\frac{t+\tau}{\tau_{ch}}\right)\right]\;.
\end{eqnarray}
For a cavity mode that is initially occupied by a coherent state with $\alpha = 1$ we show the dynamics of the fidelity for delay time $\Delta t = 14\times t_{c_2}$ in Fig.~\ref{fig:delayed_transfer}(a).  
Before the transfer the fidelity is equal to one, but after the first set of pulses is applied, the cavity empties and the fidelity drops to the value for the vacuum, $F\sim |\langle \Psi_i|0\rangle|^2$ (indicated by the orange-colored line). This means that the quantum state initially present in the cavity mode has been almost fully transferred to the mechanical mode, from which it will be retrieved when the second set of pulses is applied. At this point the cavity state fidelity increases again and in Fig.~\ref{fig:delayed_transfer}(b) the corresponding mode occupations are shown.  One can see that the cavity mode occupation goes down to zero after the initial transfer, while the mechanical mode occupation rises. The latter also slowly increases during the storage period due to its coupling to the thermal environment at $T=1\,{\rm mK}$, the steady-state mechanical occupation tending towards $\bar{n}_b\sim 1.6$. During application of the transfer pulses the magnon mode occupation rises briefly, but also quickly returns to zero. 
In Fig.~\ref{fig:delayed_transfer}(c), we show the cavity fidelity after retrieval, $F_r$ (red dots) as a function of the storage time, $t_s$, with the blue line corresponding to an exponential fitting of these results. The storage time can be calculated as the time difference between the application of the transfer and retrieval pulses, which can be given approximately as $t_s = (t_{c_2}+\Delta t - t_{c_1})$. The half-life of the fitted  exponential decay rate is given by 6 ms, where the primary loss arises from the mechanical mode damping, which itself has a lifetime of 10 ms. The additional small losses can therefore be attributed to magnon mode decay that occurs during the transfer. 


 \begin{figure}[t] 
 	\begin{center}
 		\setlength{\unitlength}{1cm}
 		\begin{picture}(8.5,16)
 		 		\put(0.5,9){\includegraphics[width=.55\columnwidth]{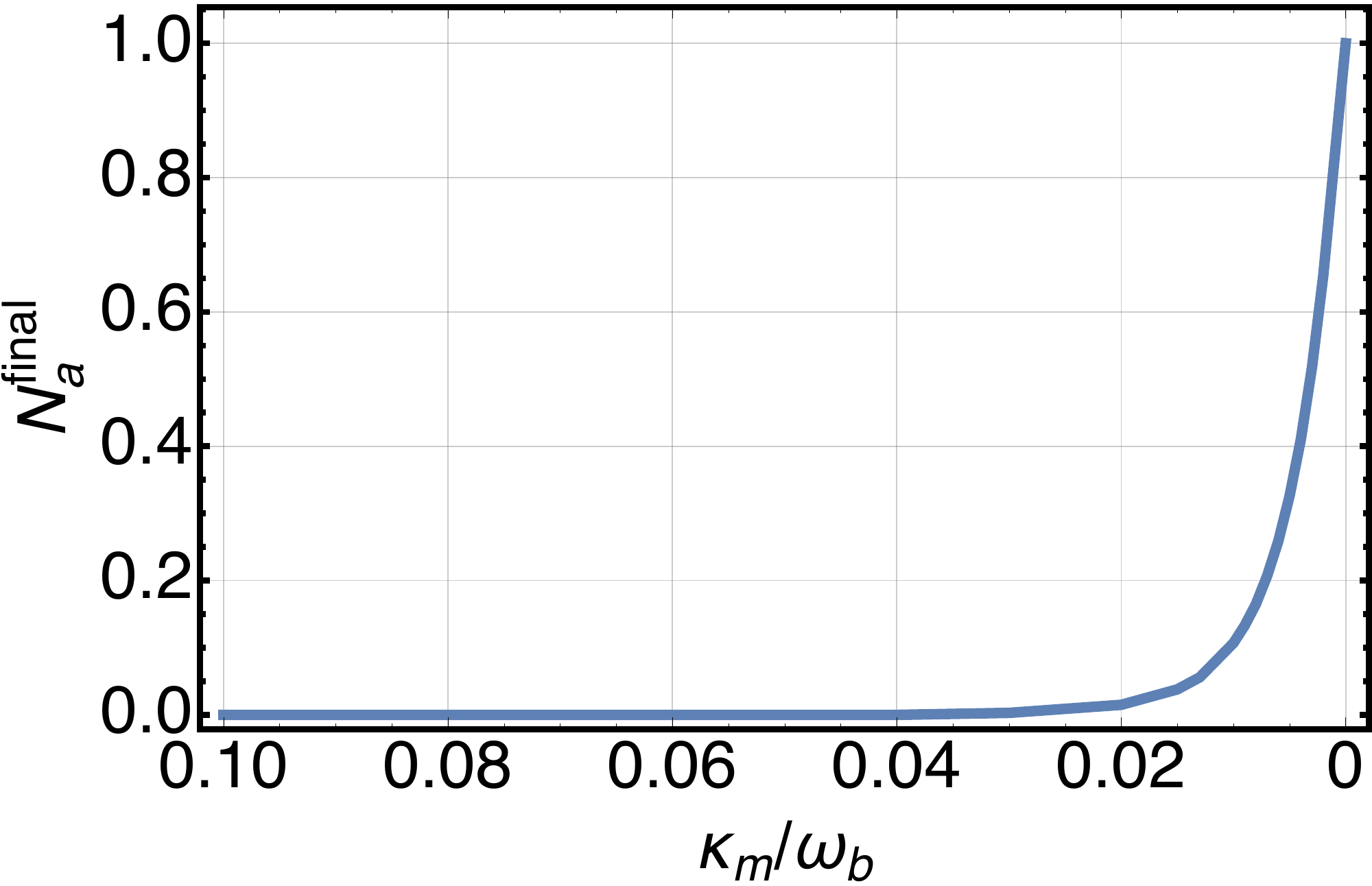}}
		\end{picture}
 	\end{center}
 	\vspace{-8.5cm}
 	\caption{The number of retrieved cavity photons, ${{N_a}^{\rm final}}$, as a function of the magnon damping rate $\kappa_m/\omega_b$ at a bath temperature of $T_{\rm th} = 10\  \rm mK$. The parameters considered here are: $2\pi \Omega_0/\omega_b=0.5$, $\omega_b T = 108.7/m$, $\omega_b t_{c_1} = -612.2/m$, $\omega_b t_{c_2} = 612.2/m$, $\omega_b \tau_{ch} = 164.9/m$, $\omega_b \tau = 1101.6/m$, $\Delta t = 0$, where $m=5$. Also $2\pi \kappa_b/\omega_b = 10^{-5}$, $\kappa_{\delta} = 14.05$, and $h_{\delta} = 13.94$. Here $\kappa_\delta$ and $h_\delta$ are dimensionless parameters.  
}
\label{fig:high_temp_kappam}
 \end{figure}
 
For all the analysis above, we have assumed rather low values for the bath temperature and the magnon damping rate, which are currently not experimentally accessible. Let us therefore in the following explore the effects of higher bath temperature and higher magnon damping. Since the numerical treatment becomes very resource intensive for higher bath occupations \cite{johansson2012qutip,johansson2012qutip2}, we first solve for the average values of the second-order moments of the system analytically as described in the following. Applying the master equation given in Eq.~(\ref{master_eq}), a linear set of differential equations for the second-order moments can be obtained as 
\begin{equation}
{\partial }_t\left\langle {\hat{o}}_i{\hat{o}}_j\right\rangle =Tr\left(\dot{\rho }{\hat{o}}_i{\hat{o}}_j\right)=\sum_{m,n}{{\mu}_{m,n}\left\langle {\hat{o}}_m{\hat{o}}_n\right\rangle }, 
\end{equation} 
where the ${\hat{o}}_i$, ${\hat{o}}_j$, ${\hat{o}}_m$, ${\hat{o}}_n$ are one of the operators: ${a}^{\dagger }_1$, $m^{\dagger }_1$, ${b}^{\dagger }_1$, $a_1$, $m_1$ and $b_1$; and ${\mu }_{m,n}$ are the corresponding coefficients. 
We use this approach to time evolve the mean occupations in the 
photon ($\langle a^{\dagger }_1 a_1 \rangle $), phonon ($\langle b^{\dagger }_1 b_1\rangle $) and magnon ($\langle m^{\dagger }_1 m_1\rangle $) modes,
considering that initially only the cavity mode is occupied with $\langle N_a\rangle (= \langle a^{\dagger }_1 a_1\rangle) (t=0) = 1$, and all the other second-order moments are zero, with state $|N_a, N_m, N_b \rangle (t=0) = |1, 0, 0\rangle$. Considering $T_{\rm th} = 10\  \rm mK$, we show in Fig.~\ref{fig:high_temp_kappam} the retrieved photon  number in the cavity at the end of the scheme as a function of a decreasing magnon damping rate. Here the storage time is chosen to be $\omega_b t_s = 1224.4$ which for $\omega_b/2\pi = 10\  \rm MHz$ corresponds to $t_s = 0.19 \ \rm ms$.  One can see that, as expected, lower magnon damping rates result in improved photon retrieval.

\section{Conclusions}
In conclusion, we have presented a scheme to transfer quantum states from a microwave cavity mode to a mechanical mode in a photon-magnon-phonon hybrid system where a YIG sphere is placed in a microwave cavity with the magnon mode coupled both to the photonic and phononic modes. We have shown that using time-modulated shapes for the detunings for the magnon and cavity modes and also the magnetostrictive coupling, transfer with high fidelity between non-directly coupled modes can be obtained for a number of different states including coherent states, cat states and squeezed vacuum states. 
Given the flexibilities in controlling the magnomechanical coupling, it is interesting to consider this work as a first step for implementing mechanical bosonic quantum error correction codes on the stored phononic quantum information to further increase the mechanical storage times \cite{terhal2020towards}.

\section*{Acknowledgements}
This work was supported by the Okinawa Institute of Science and Technology Graduate University. We are grateful for the help and support provided by the Scientific Computing and Data Analysis section of Research Support Division at OIST. We acknowledge support from the ARC Centre of Excellence for Engineered Quantum Systems grant CE170100009.

\section*{References}
\bibliographystyle{iopart-num}
\bibliography{references}

\providecommand{\newblock}{}
\begin{thebibliography}{10}
\expandafter\ifx\csname url\endcsname\relax
  \def\url#1{{\tt #1}}\fi
\expandafter\ifx\csname urlprefix\endcsname\relax\def\urlprefix{URL }\fi
\providecommand{\eprint}[2][]{\url{#2}}

\bibitem{kippenberg2005analysis}
Kippenberg T, Rokhsari H, Carmon T, Scherer A and Vahala K 2005 {\em Phys. Rev.
  Lett.\/} {\bf 95} 033901

\bibitem{arcizet2006radiation}
Arcizet O, Cohadon P~F, Briant T, Pinard M and Heidmann A 2006 {\em Nature\/}
  {\bf 444} 71--74

\bibitem{teufel2011circuit}
Teufel J~D, Li D, Allman M, Cicak K, Sirois A, Whittaker J and Simmonds R 2011
  {\em Nature\/} {\bf 471} 204--208

\bibitem{andrews2014bidirectional}
Andrews R~W, Peterson R~W, Purdy T~P, Cicak K, Simmonds R~W, Regal C~A and
  Lehnert K~W 2014 {\em Nature Phys.\/} {\bf 10} 321--326

\bibitem{wang2012using}
Wang Y~D and Clerk A~A 2012 {\em Phys. Rev. Lett.\/} {\bf 108} 153603

\bibitem{tian2012adiabatic}
Tian L 2012 {\em Phys. Rev. Lett.\/} {\bf 108} 153604

\bibitem{hill2012coherent}
Hill J~T, Safavi-Naeini A~H, Chan J and Painter O 2012 {\em Nature Comm.\/}
  {\bf 3} 1--7

\bibitem{Zhang2016a}
Zhang X, Zou C~L, Jiang L and Tang H~X 2016 {\em Sci. Adv.\/} {\bf 2} e1501286

\bibitem{serga2010yig}
Serga A, Chumak A and Hillebrands B 2010 {\em J. Phys. D: Appl. Phys.\/} {\bf
  43} 264002

\bibitem{lenk2011building}
Lenk B, Ulrichs H, Garbs F and M{\"u}nzenberg M 2011 {\em Phys. Rep.\/} {\bf
  507} 107--136

\bibitem{chumak2015magnon}
Chumak A~V, Vasyuchka V~I, Serga A~A and Hillebrands B 2015 {\em Nature
  Phys.\/} {\bf 11} 453--461

\bibitem{Kittel1948b}
Kittel C 1948 {\em Phys. Rev.\/} {\bf 73} 155--161

\bibitem{Tabuchi2015}
Tabuchi Y, Ishino S, Noguchi A, Ishikawa T, Yamazaki R, Usami K and Nakamura Y
  2015 {\em Science\/} {\bf 349} 405--408

\bibitem{Huebl2013}
Huebl H, Zollitsch C~W, Lotze J, Hocke F, Greifenstein M, Marx A, Gross R and
  Goennenwein S~T 2013 {\em Phys. Rev. Lett.\/} {\bf 111} 127003

\bibitem{Tabuchi2014}
Tabuchi Y, Ishino S, Ishikawa T, Yamazaki R, Usami K and Nakamura Y 2014 {\em
  Phys. Rev. Lett.\/} {\bf 113} 083603

\bibitem{Zhang2014}
Zhang X, Zou C~L, Jiang L and Tang H~X 2014 {\em Phys. Rev. Lett.\/} {\bf 113}
  156401

\bibitem{Goryachev2014}
Goryachev M, Farr W~G, Creedon D~L, Fan Y, Kostylev M and Tobar M~E 2014 {\em
  Phys. Rev. Appl.\/} {\bf 2} 054002

\bibitem{Bai2015}
Bai L, Harder M, Chen Y~P, Fan X, Xiao J~Q and Hu C~M 2015 {\em Phys. Rev.
  Lett.\/} {\bf 114} 227201

\bibitem{Zhang2015a}
Zhang D, Wang X~M, Li T~F, Luo X~Q, Wu W, Nori F and You J 2015 {\em npj
  Quantum Inf.\/} {\bf 1} 1

\bibitem{li2018magnon}
Li J, Zhu S~Y and Agarwal G 2018 {\em Phys. Rev. Lett.\/} {\bf 121} 203601

\bibitem{li2019squeezed}
Li J, Zhu S~Y and Agarwal G 2019 {\em Phys. Rev. A\/} {\bf 99} 021801

\bibitem{li2021entangling}
Li J and Gr{\"o}blacher S 2021 {\em Quant. Sci. Tech.\/} {\bf 6} 024005

\bibitem{qi2020magnon}
Qi S~F and Jing J 2020 {\em arXiv:2011.05642\/}

\bibitem{Wang2018}
Wang Y~P, Zhang G~Q, Zhang D, Li T~F, Hu C~M and You J~Q 2018 {\em Phys. Rev.
  Lett.\/} {\bf 120} 057202

\bibitem{briegel1998quantum}
Briegel H~J, D{\"u}r W, Cirac J~I and Zoller P 1998 {\em Phys. Rev. Lett.\/}
  {\bf 81} 5932

\bibitem{liu2001observation}
Liu C, Dutton Z, Behroozi C~H and Hau L~V 2001 {\em Nature\/} {\bf 409}
  490--493

\bibitem{phillips2001storage}
Phillips D~F, Fleischhauer A, Mair A, Walsworth R~L and Lukin M~D 2001 {\em
  Phys. Rev. Lett.\/} {\bf 86} 783

\bibitem{hammerer2010quantum}
Hammerer K, S{\o}rensen A~S and Polzik E~S 2010 {\em Rev. Mod. Phys.\/} {\bf
  82} 1041

\bibitem{fiore2011storing}
Fiore V, Yang Y, Kuzyk M~C, Barbour R, Tian L and Wang H 2011 {\em Phys. Rev.
  Lett.\/} {\bf 107} 133601

\bibitem{fiore2013optomechanical}
Fiore V, Dong C, Kuzyk M~C and Wang H 2013 {\em Phys. Rev. A\/} {\bf 87} 023812

\bibitem{kumar2019single}
Kumar P and Bhattacharya M 2019 {\em Phys. Rev. A\/} {\bf 99} 023811

\bibitem{yanik2004stopping}
Yanik M~F and Fan S 2004 {\em Phys. Rev. Lett.\/} {\bf 92} 083901

\bibitem{xu2007breaking}
Xu Q, Dong P and Lipson M 2007 {\em Nature Phys.\/} {\bf 3} 406--410

\bibitem{baba2008slow}
Baba T 2008 {\em Nature Photon.\/} {\bf 2} 465--473

\bibitem{zhu2007stored}
Zhu Z, Gauthier D~J and Boyd R~W 2007 {\em Science\/} {\bf 318} 1748--1750

\bibitem{bergmann2019roadmap}
Bergmann K, N{\"a}gerl H~C, Panda C, Gabrielse G, Miloglyadov E, Quack M,
  Seyfang G, Wichmann G, Ospelkaus S, Kuhn A {\em et~al.\/} 2019 {\em J. Phys.
  B: At. Mol. Opt. Phys.\/} {\bf 52} 202001

\bibitem{bergmann2015perspective}
Bergmann K, Vitanov N~V and Shore B~W 2015 {\em J. Chem. Phys.\/} {\bf 142}
  170901

\bibitem{carvalho2014piezoelectric}
Carvalho N~C, Fan Y, Le~Floch J~M and Tobar M~E 2014 {\em Rev. Sci. Instrum.\/}
  {\bf 85} 104705

\bibitem{c2016piezoelectric}
C~Carvalho N, Fan Y and Tobar M 2016 {\em Rev. Sci. Instrum.\/} {\bf 87} 094702

\bibitem{clark2018cryogenic}
Clark T, Vadakkumbatt V, Souris F, Ramp H and Davis J 2018 {\em Rev. Sci.
  Instrum.\/} {\bf 89} 114704

\bibitem{ramp2020wavelength}
Ramp H, Clark T, Hauer B, Doolin C, Balram K~C, Srinivasan K and Davis J 2020
  {\em Appl. Phys. Lett.\/} {\bf 116} 174005

\bibitem{vitanov2017stimulated}
Vitanov N~V, Rangelov A~A, Shore B~W and Bergmann K 2017 {\em Rev. Mod.
  Phys.\/} {\bf 89} 015006

\bibitem{nielsen2011quantum}
Nielsen M~A and Chuang I~L 2011 {\em Quantum Computation and Quantum
  Information\/} (Taylor \& Francis)

\bibitem{teufel2011sideband}
Teufel J~D, Donner T, Li D, Harlow J~W, Allman M, Cicak K, Sirois A~J,
  Whittaker J~D, Lehnert K~W and Simmonds R~W 2011 {\em Nature\/} {\bf 475}
  359--363

\bibitem{schliesser2006radiation}
Schliesser A, Del’Haye P, Nooshi N, Vahala K and Kippenberg T~J 2006 {\em
  Phys. Rev. Lett.\/} {\bf 97} 243905

\bibitem{sarma2020optomechanical}
Sarma B, Busch T and Twamley J 2020 {\em New J. Phys.\/} {\bf 22} 103043

\bibitem{johansson2012qutip}
Johansson J~R, Nation P~D and Nori F 2012 {\em Comput. Phys. Commun.\/} {\bf
  183} 1760--1772

\bibitem{johansson2012qutip2}
Johansson J, Nation P and Nori F 2013 {\em Comput. Phys. Commun.\/} {\bf 184}
  1234--1240

\bibitem{terhal2020towards}
Terhal B, Conrad J and Vuillot C 2020 {\em Quant. Sci. Tech.\/} {\bf 5} 043001

\end{thebibliography}
\end{document}